\documentclass[conference]{IEEEtran}
\makeatletter
\def\ps@headings{%
\def\@oddhead{\mbox{}\scriptsize\rightmark \hfil \thepage}%
\def\@evenhead{\scriptsize\thepage \hfil \leftmark\mbox{}}%
\def\@oddfoot{}%
\def\@evenfoot{}}
\makeatother
\usepackage{amssymb,amsmath,amsthm}
\usepackage{graphicx}
 \usepackage{caption}
\usepackage{subcaption}
\usepackage{mdwmath}
\usepackage{mdwtab}
\usepackage{soul}
\usepackage{cite}
 \usepackage{multirow}
 \usepackage{color} 
  \usepackage[linesnumbered,ruled]{algorithm2e}
\usepackage{setspace}
\usepackage[hyphenbreaks]{breakurl}
\newtheorem{definition}{Definition}

\begin{document}
   \title{Optimized Parallel Transmission in Elastic Optical Networks to Support High-Speed Ethernet}
   \author{\IEEEauthorblockN{Xiaomin Chen$^1$, Admela Jukan$^1$, Ashwin Gumaste$^2$}
\IEEEauthorblockA{Technische Universit\"at Carolo-Wilhelmina zu Braunschweig, Germany$^1$\\
Department of Computer Science and Engineering, Indian Institute of Technology, Bombay, India$^2$}
Email: \{chen, jukan\}@ida.ing.tu-bs.de, \{ashwing\}@ieee.org }
\maketitle

\begin{abstract}  
The need for \emph{optical parallelization} is driven by the imminent \emph{optical capacity crunch}, where the spectral efficiency required in the coming decades will be beyond the Shannon 
limit. To this end,  the emerging high-speed Ethernet services at 100 Gbps, have already standardized options to utilize 
parallel optics to parallelize interfaces referred to as \emph{Multi-lane Distribution} (MLD). OFDM-based optical 
network is a promising transmission option towards the goal of Ethernet parallelization. It can allocate optical 
resource tailored for a variety of bandwidth requirements and that in a fundamentally parallel fashion with each 
sub-carrier utilizing a frequency slot at a lower rate than if serial transmission was used. In this paper, we propose a novel 
parallel transmission framework designed for elastic (OFDM-based) optical networks to support high-speed Ethernet 
services, in-line with  IEEE and ITU-T standards.  We formulate an ILP optimization model based on integer linear 
programming, with consideration of various constraints, including spectrum fragmentation, differential delay and 
guard-band constraints. We also propose a heuristic algorithm which can be applied when the optimization model becomes 
intractable. The numerical results show the effectiveness and high suitability of elastic optical networks to support 
parallel transmission in high-speed Ethernet.  To the best of our knowledge, this is  the first attempt to investigate the parallel 
transmission in elastic optical networks to support standardized high-speed Ethernet.

 \end{abstract}

\section{Introduction}     
\par The need for \emph{optical parallelization} is driven by the imminent \emph{optical capacity crunch}~\cite{Chralyvy:ecoc:2009}, which has gained significant attention after studies have shown that capacity upgrades of conventional single-mode fiber systems have slowed down from about $80\%$ per year to about $20\%$ per year since 2002~\cite{Chralyvy:ecoc:2009}\cite{Winzer:2012}.  Given the massive growth of Internet data, it became clear that optical communications had to shift towards spectral efficiency, i.e., transmitting more information over the fundamentally limited bandwidth of optical amplifiers. It has become apparent that without parallelization, the spectral efficiency of about 20b/s/Hz required in the coming decades, will be beyond the Shannon limit \cite{Winzer:2012}. As a consequence, the high-speed Ethernet has resorted to a parallel solution as standardized in IEEE802.3ba,  especially for 40/100 Gigabit Ethernet (40/100GE) and beyond, to overcome the Shannon limit without 
an expense on reduced transmission distance. Instead of using  high-speed serial interfaces, 40GE and 100GE   utilize 
parallel optics to split traffic across multiple lanes with lower rates, which is referred to as \emph{Multi-lane 
Distribution} (MLD) \cite{802.3ba}. To this end, it has been specified that 40 Gbps Ethernet (40GE) and 100 Gbps 
Ethernet (100GE) can utilize 4 lanes and 10 lanes,  respectively, with each lane running at 10.3125 Gbps 
\cite{802.3ba}. To this end,  ITU-T has extended the Optical Transport Network (OTN) information structure especially 
for 40GE and 100GE traffic.  As it is, the concept of parallel transmission for high speed Ethernet presents  a myriad of 
new challenges in the optical layer. 

\par We believe that recently proposed elastic optical networks based on Orthogonal Frequency Division Multiplexing (OFDM)  technology carry the promise of a potentially transformative parallel transmission solution in the optical layer, due to its fundamental parallel nature.   With OFDM, optical spectrum is  actually \emph{sliced or parallelized}  into a sequence of frequency slots and  signals are modulated on frequency slots in form of sub-carriers. Since the sub-carriers are orthogonal in the frequency domain, signals modulated on them can be received in parallel without interference. Hence, the OFDM-based elastic optical networks can efficiently support  parallel transmission of high-speed Ethernet by modulating  traffic on a group of parallel sub-carriers,  with each sub-carrier at a lower data rate.  High-speed Ethernet can thus take full advantage of Optical Virtual Concatenation (OVC) protocol in OTNs for parallel transmission over  spectrum bands in OFDM networks, where a spectrum band is composed of a group of consecutive sub-carriers. With OVC protocol,  $n$ Ethernet lanes are  carried by $n$ sub-carriers in multiple spectrum bands as though they were virtually contiguous. At the same time,  parallel transmission can easily alleviate the fundamental trade-off between bit rate and the optical reach, as it is known that all-optical transmission distance falls as the serial bit rate increases. This is due to the imperfections in the optical fibers, such as transmission loss, non-linear effects, group velocity dispersion (GVD), polarization mode dispersion (PMD), etc.

\begin{figure*}[ht] 
	\centering
		\includegraphics[height= 11.5cm, width=1.92\columnwidth]{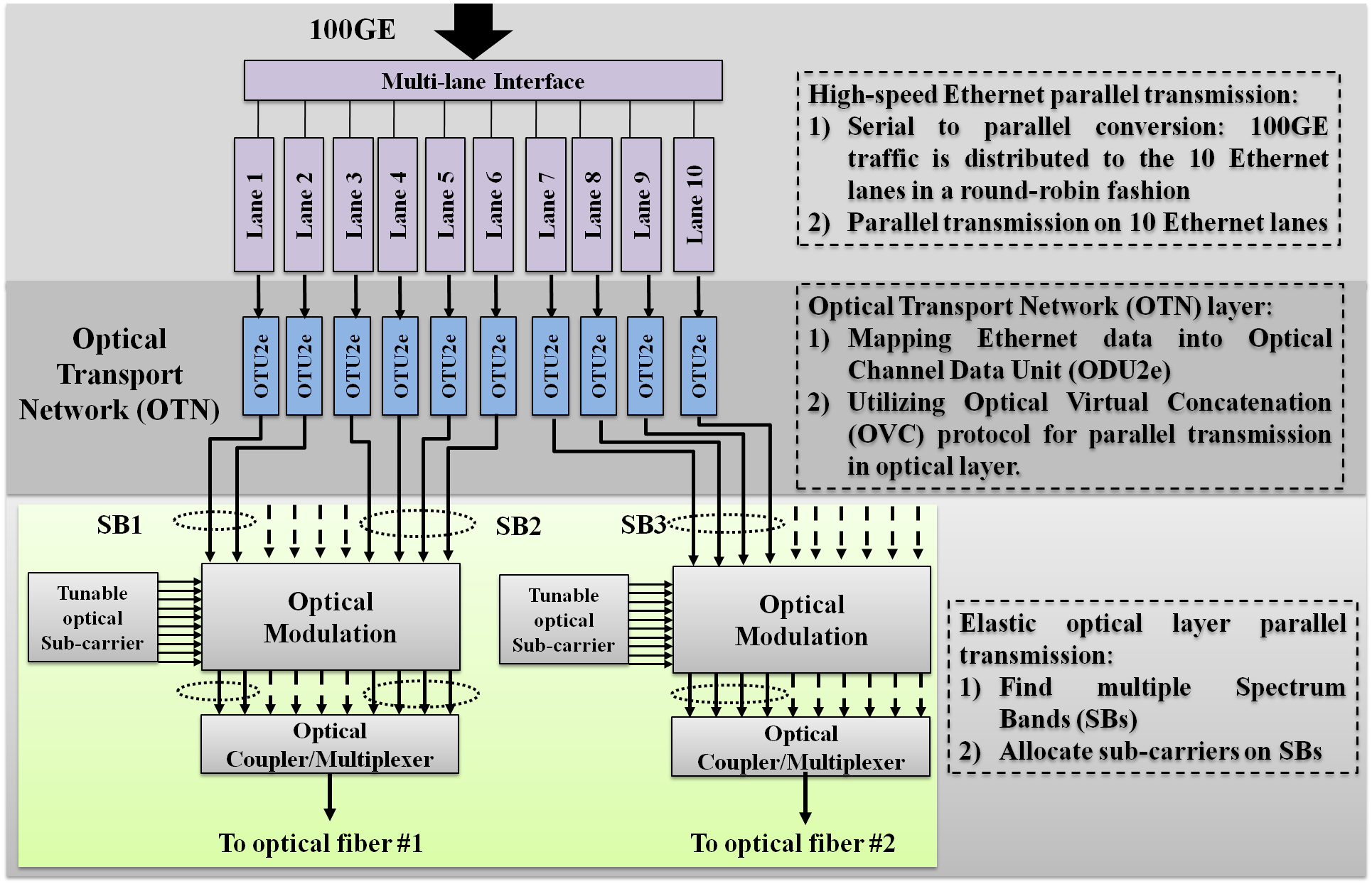}
	\caption{Parallel transmission in elastic optical networks to support high-speed Ethernet.} 
	\label{fig:imple}
	\vspace{-0.3cm}
\end{figure*}   
\par We present an illustrative example in Fig.\ref{fig:imple}, where spectrum on each fiber is \emph{sliced} or, as we also  refer it to  as \emph{parallelized} into 10 sub-carriers (dashed lines are the subcarriers already "allocated"). OTN layer acts as an adaptation  layer  mapping the Ethernet traffic onto optical frames (ODUs).  Here, 100Gbps Ethernet 
 (100GE) signal  is firstly distributed into 10  lanes in a round-robin fashion in the Ethernet layer, each lane carrying a 
 10Gbps Ethernet signal. After that, each lane is mapped onto Optical channel Data Unit (ODU) of appropriate size, 
 i.e., ODU2e, which are then mapped into Optical channel Transport Unit (OTU2e) at speed of approximately 11.09Gbps. In the 
 optical layer, data carried on each  OTU2e channel is modulated to optical signal and transmitted on sub-carriers in the 
 elastic optical network.  The number of sub-carriers depends  on the   high-speed Ethernet transmission rate and  capacity per sub-carrier.  Assume  each sub-carrier can support one OTU2e channel, 10 sub-carriers are required   in this example. We define a group of concecutive sub-carriers as a \emph{Spectrum Band} (SB). A connection may be set up using multiple spectrum bands, depending on the spectrum availability. For instance, three spectrum bands are used in Fig.\ref{fig:imple}, i.e., SB1, SB2 and SB3.

\par In this paper, we investigate the suitability of OFDM-based networks to support parallel transmission in high speed 
Ethernet, and address the imminent challenges in its implementation. First, we address the issue \emph{spectrum 
fragmentation} ~\cite{Chen:infocom:2013}\cite{Chen:ICC:2013}.  Most of Routing and Spectrum Assignment 
(RSA) methods focu on allocating sub-carriers required by a connection in a consecutive spectrum range, which can not take full advantage of parallel nature of elastic optical networks. We show that optimization methods can be used to 
minimize the  spectrum fragmentation. Second, we show that also the \emph{differential delay issue} can be 
effectively addressed, as it normally occurs when traffic is distributed into sub-carriers across diverse fibers. 
Specifically, we not only optimize the differential delay, but we also show that it can be bounded according to the 
requirements in the current standards and commercial products\footnote{ITU-T G.709 \cite{ITU-T:G.709} suggests 
that the realignment process has to be able to compensate a differential delay of at least $\pm$125$\mu s$. In terms of 
commercial products, it has been reported that a commercial framer device can support $250 \mu s$ differential delay 
using internal memory and up to $128ms$ using off-chip Synchronous Dynamic Random Access Memory 
(SDRAM)\cite{cisco}\cite{Intel:datasheet}.}.  We formulate a novel optimization model based on Integer Linear 
Programming (ILP) to find an optimal set of sub-carriers which are used in the parallel transmission, while minimizing 
overall usage of optical spectrum. In addition, we propose a heuristic algorithm which can be applied in the scenarios  
where the optimization model becomes infeasible.    The numerical results show the effectiveness and high suitability 
of elastic optical networks to support parallel transmission in high-speed Ethernet.  To the best of our knowledge, this 
is the first attempt to investigate the parallel transmission in elastic optical networks to support standardized high-
speed Ethernet.

\par The rest of the paper is organized as follows. Section \ref{implementation} present a background overview, 
 including the differential delay issue, spectrum fragmentation and  key  technologies  as well as the implications of the assumptions we make. In Section \ref{algo}, we present the ILP model  and a heuristic algorithm. We show the numerical results in  Section \ref{performance}. Section \ref{relatedwork} provides a literature review.   We conclude the paper in  Section \ref{conclusion}.

\section{Background}\label{implementation}
\subsection{Key Technologies} 
\par The feasibility of parallel transmission problem studied in this paper stems from the inherent  parallelism in the high-speed Ethernet and elastic OFDM-based optical layers, as specified in the current standards.  Traditionally,  Gigabit Ethernet transmission relies on duplex fiber cabling with one fiber deployed for each direction. The high-speed Ethernet standard, i.e., IEEE 802.3ba, specified Multiple Lane Distribution (MLD) to  split   high-speed serial Ethernet data  into multiple virtual lanes in a round-robin fashion, as it was shown earlier in  Fig.\ref{fig:imple}. As a result, the parallel optics systems are used in every Ethernet cabling solution. For instance, the 40Gbps  Ethernet calls for a solution of  12-fiber cabling, with  each channel featuring four dedicated fibers for transmitting and receiving respectively, while saving  the remaining  four fibers as dark fibers for reliability~\cite{802.3ba} .   
  
\par In the optical layer, the built-in inverse multiplexing protocol in OTN, i.e.,  optical virtual concatenation (OVC), has been proposed to enable parallel transmission in optical networks precisely.  As standardized in ITU-T G.709, OTN defines a set of \emph{optical wrappers} with different size  \cite{Gumaste:Comag:2010}. The client traffic is  mapped  into an optical wrapper of appropriate size and transmitted at corresponding line rates, referred to as OTU$k$. The OTN information structure supports  line rates varying from approximately 2.5Gbps ($k=1$) to 112 Gbps ($k=4$).  Of note is that the OTN information structure  has been recently updated in order to support the new emerging high-speed Ethernet services, in particular for 40Gbps and 100Gbps Ethernet. Two new optical wrappers are defined, referred to as, OTU3e2 (44.58Gbps) and OTU4 (112Gbps) respectively~\cite{ITU-T:G.709}.   The technological maturity of OTN layer, including synchronization, error correction, framing  and differential delay compensation, can be utilized to facilitate the parallel transmission in elastic optical networks to support high-speed Ethernet.  Moreover, elastic optical networks with OFDM technologies enables the mapping between the optical wrappers and sub-carriers enabling parallel transmission without complicated data processing.

\subsection{Differential Delay in Parallel Transmission in  OFDM-based Optical Networks}  
\par In general, parallel transmission in OFDM-based networks falls  under two categories: (i) single spectrum band transmission, (ii) multiple spectrum bands transmission. In the first scenario,  the required number of sub-carriers can be allocated within a single spectrum band from source to destination. In the second scenario, multiple spectrum bands are allocated to support the high-speed Ethernet transmission. The multiple spectrum bands can be transmitted over the same or diverse fibers from source to destination. In either category, each sub-carrier generally experiences different end-to-end delay, i.e., even when transmitted over the same fibers and within the same spectrum band. For instance, in Fig.\ref{fig:imple}, the sub-carriers in spectrum bands $SB1$ and $SB2$ can experience the delay difference resulting from the imperfections in optical fibers, even if transmitted over the same optical path or different paths with the same length. Thus, we distinguish between two main types of differential delay in parallel transmission in OFDM-based networks: (i) fiber effects caused differential delay, and (ii) path diversity caused differential delay.

\subsubsection{Fiber effects caused differential delay} 
The main fiber effect to cause differential delay is the Group Velocity Dispersion (GVD), which is caused by the fact that sub-carriers on different frequencies travel (in form of waves) at different speeds. The longer wave travels faster than the shorter waves, resulting in different propagation delay even in the same spectrum band.  According to the principle of GVD, the maximum delay difference caused by GVD in a spectrum band can be approximated as: 
\begin{equation} \label{GVD}  
 \Delta  d_{max} \approx  D(f_c) \cdot (f_{max}-f_{min}) \cdot L
\end{equation}
where $D(f_c)$ is the fiber dispersion at the central frequency; $f_{max}$ and $f_{min}$ are the highest frequency and smallest frequency of the spectrum band, see~\cite{Sun:2008}; and $L$ is the transmission distance.
The condition of Eq.\eqref{GVD} is that the central frequency is much larger than $(f_{max}-f_{min})$. The OFDM-based optical networks follow the same spectrum dimension as it is in  \emph{``fix grid"} \cite{Jinno:2009}, i.e., the central frequency is $f_c$ = 193.1 THz, which is much larger than the frequency difference value   in any spectrum band. Hence, the approximation presented in Eq.\eqref{GVD} can be directly applied. As an example,  we assume that 100GE utilizes a spectrum band composed of 10 consecutive sub-carriers for parallel transmission and also assume that each frequency slot is 50GHz\footnote{A frequency slot is generally smaller than 50GHz. Here, we assume the standardized channel spacing, i.e., 50GHz as an example. }, i.e., channel spacing is 
0.4nm. The fiber dispersion of a Single Mode Fiber (SMF) is 17 ps/nm/km at the the central frequency \cite{Sun:2008}. Hence, the maximum differential delay caused by dispersion in the parallel transmission is  $\Delta d{max} \approx$ 0.68 $\mu s$ for a connection with physical distance of $1 \times 10^4 km$. 

\subsubsection{Path diversity caused differential delay} 
Another, and more commonly considered, type of differential delay in parallel transmission is caused by the path diversity. When spectrum bands used in the parallel transmission traverse different fiber-level paths,  the different transmission distance along different paths also lead to the differential delay. The differential delay caused by the path diversity can be simply calculated as $L/v$, where $v$ is the signal propagation speed in the path and $L$ is the path length. Considering the standard SMF where the signal propagation  speed is $2 \times 10^5 km/s$, for a  connection with length of $1 \times 10^4 km$,  the  maximum differential delay between spectrum bands would be $50ms$.

 \subsection{Spectrum Fragmentation}
 \begin{figure*}[ht]
	\centering
		\includegraphics[width=1.6\columnwidth]{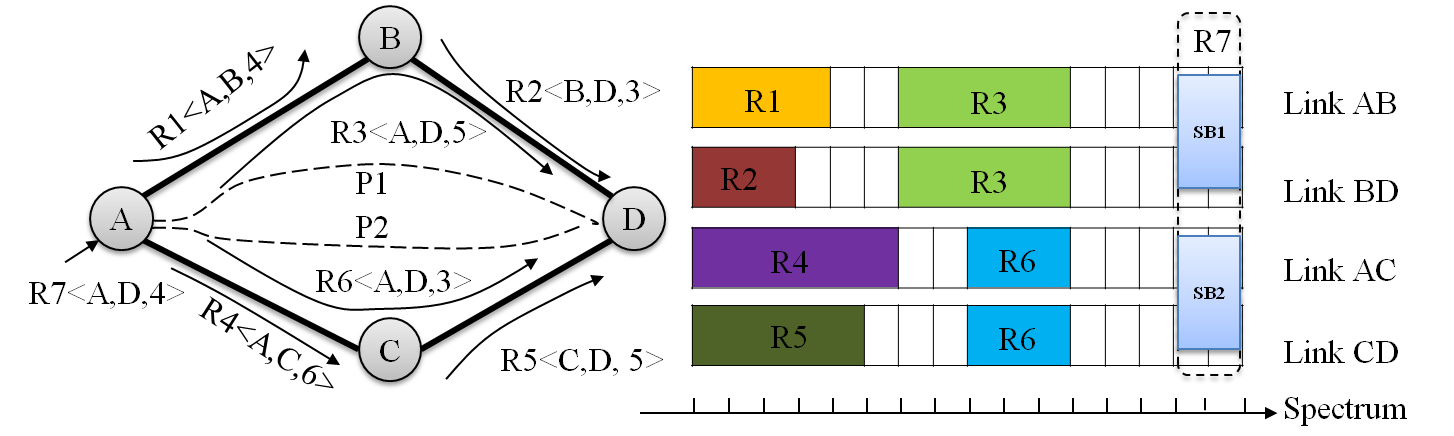}
	\caption{An illustrative example of spectrum fragmentation and a solution with two non-consequitve spectrum slices} \vspace{-0.3cm}	\label{fig:spectrumFrag}
\end{figure*}
\par As previously mentioned, the nature of connection demands from high-speed Ethernet, which can be fine and 
coarse grained, may exacerbate the so-called \emph{spectrum fragmentation}. To better understand this problem,  we 
show an  example  in Fig.\ref{fig:spectrumFrag}, where optical spectrum on each fiber link is assumed to be  sliced 
into 16 frequency slots, with one sub-carrier on each frequency slot.  The traffic demand is defined as  $R(S, D,T_r)$, 
where $S$, $D$ and $T_r$ denote source, destination, and the number of required  sub-carriers, respectively. In this 
example, 2 sub-carriers are assigned as the guard-band (typically used to insulate the adjacent spectrum bands).  As it 
can be seen in Fig.\ref{fig:spectrumFrag}, the spectrum  on all fiber links are  fragmented after the allocation of 6 
spectrum bands, i.e., $R_1-R_6$.  
 Upon the arrival of $R_7$, the   request is rejected  if it is only allowed to set up a single spectrum path with four consecutive 
 sub-carriers. In fact, under current network condition, any traffic demand requesting more than 3 sub-carriers will be 
 blocked, even  though there are sufficient sub-carriers available in the network. This phenomena is particularly 
 pronounced in case of  high-speed Ethernet services  with a transmission rate of   40/100 Gbps which require  a large 
 number of consecutive sub-carriers. However, this issue can be effectively resolved by  distributing Ethernet 
 traffic into non-consecutive parallel spectrum paths.    In the example shown in Fig.\ref{fig:spectrumFrag}, $R_7$ can 
 be accommodated by allocating two spectrum bands, i.e., $SB_1$ and $SB_2$, with two sub-carriers per band.

 \subsection{Assumptions and Discussion}
\par In this paper we assume all sub-carriers have the same transmission rate. In OFDM-based optical networks,  however, it has 
been shown that the capacity per sub-carrier can be adaptively managed by using different modulation formats for 
different transmission distances~\cite{Christo:JLT:2011}\cite{Jinno:2010}. For 
instance, in our Fig.\ref{fig:imple}, spectrum band 1 (SB1) would use a transmitter at higher transmission speed with a higher level modulation format. In this paper, however, we stay with  the assumption that all sub-carriers have 
the same transmission rate for practical implementation.  This assumption  simplifies the  mapping between OTN frames and sub-carriers, thus  facilitating the actual implementation of optical parallel transmission in practice. 
\par The analysis of differential delay issue shown earlier has indicated that the main factor of differential delay in parallel transmission is path diversity. Considering the same transmission distance, the maximum differential delay caused by the GVD among sub-carriers is insignificant comparing with the delay difference caused by propagation, e.g., $0.68\mu s$ vs. $50ms$.  Therefore, our results  for heuristic evaluation will only consider the differential delay caused by path diversity. However, the optimization model proposed in the next section can account for the issue of differential delay among  sub-carriers, which maybe useful for new materials inducing new   fiber propagation properties. Finally, we assume that the differential delay is compensated in the OTN layer in the our architecture.

\section{Parallel Transmission Algorithms}\label{algo}
In this section, we first present the preliminaries and provide definitions of basic concepts used. After that, we present the optimization model based on Integer Linear Programming (ILP) and a heuristic algorithm which can find single or multiple spectrum paths for parallel transmission upon the arrival of a connection 
request from high-speed Ethernet.    
\subsection{Preliminaries}
For an easy understanding of the proposed algorithms, we first summarize the notations in Table \ref{tab:notation} and clarify the terminologies as follows:
\begin{itemize}
\item Sub-carrier:  Alike wavelength in WDM networks, a sub-carrier is a channel  which carries signals in OFDM based elastic optical networks. Sub-carriers are orthogonal to each other; spacing between their central frequencies is $nf$, where $f$ is a frequency slot.
\cite{Shieh:book}
\item Guard-band (GB): We define a guard band is the spectrum necessary to insulate two spectrum bands.
\item Spectrum path:  A spectrum path is composed of a spectrum band continuous from source to destination, i.e., it is composed of a group of consecutive sub-carriers with same index from the source to the destination node.
\item Fiber-level path: A fiber-level path is a route from source to destination, including all the fiber links in between, over which one or multiple spectrum bands can be allocated in either non-consecutive fashion. 
\end{itemize}
The terms that are specially defined for parallel transmission scenarios are given as follows:
\begin{definition}
\emph{Spectrum band (SB):} We define the \emph{Spectrum Band}  as a group of  spectrally consecutive sub-carriers allocated to the same spectrum path. 
A spectrum path can not take two spectrum bands on the same link.
\end{definition}
 \begin{definition}
\emph{Maximum acceptable differential delay:} The maximum differential delay between two spectrum paths used in the parallel transmission. It depends on the  electronic layers, e.g., OTN layer in our case, where the electronic buffer is available. It is denoted as $M$ in this paper. 
 \end{definition}
  
  \begin{table} 
\small
\begin{tabular}{|p{0.17\columnwidth}|p{0.72\columnwidth}|} 
\hline
Symbol & Description \\ \hline \hline
\multirow{2}{*}{$G(V, E)$} & A graph represents an elastic optical network with  nodes in  set $V$ and edges in set $E$ \\
$f_i$ & A sub-carrier with index $i$ \\
 $s_f$ & The size of a frequency slot\\
$F$ & An ordered set contains all frequency slots (sub-carriers) on a link, $F =\{f_1, f_2, ..., f_N\}$ \\
$F^e$ & A set  contains available frequency slots on link $e$ \\
${LD}_e$ & Delay of the link $e$,  integer value \\
$L_e$ & Length of link $e$\\
$R(S, D,T_r)$ & A connection demand, where    $S$, $D$ and $T_r$ are source, destination, the required number of sub-carriers, respectively  \\
GB &  One or multiple sub-carriers assigned to insulate two adjacent spectrum paths\\
$M$ & Maximum acceptable differential delay  \\
$p$ & A spectrum path composed of a group of consecutive sub-carriers\\
$fp$ & A fiber-level path which is a sequence of fiber links between source and destination\\
  $\mathcal{P}$ & A path set that contains all the spectrum paths computed for the connection $R$ \\ 
  $\mathcal{FP}$ & A path set that contains all the fiber-level paths computed for connection $R$ \\  
    $K$ & Maximum number of fiber-level paths can be used, $|\mathcal{FP}|\leq K$ \\  
     \hline 
\end{tabular}
\caption{Notations}
\label{tab:notation}
\end{table}

\begin{table} 
\small
\begin{tabular}{|p{0.12\columnwidth}|p{0.78\columnwidth}|} \hline
Symbol & Description\\ \hline \hline
\multirow{2}{*} {$x_{p}$} & Binary variable; it equals to 1 if a spectrum path $p $ is  used    by the connection demand, otherwise it is 0 \\
  $x_{p,e}$ & Binary variable; it equals to 1 if  a spectrum path $p$ uses $e$,   otherwise it equals to 0 \\
 $y_{p,i}$ &Binary variable; it equals to 1 if a spectrum path $p$ uses sub-carrier  $f_i \in F$, otherwise it equals 0 \\
 $x_{p,e,i}$ & Binary variable; it is 1 if a spectrum path $p$ uses sub-carrier $f_i \in F$ on link $e\in E$, otherwise it is 0 \\
 $o_{p, p'}$ & Binary variable; it equals to 1 if  spectrum paths $p$ and $p'$ share at least one link, otherwise it is  0\\ \hline
 \hline
  $pd_{p}$ & Integer variable; it denotes  delay of   spectrum path $p$ \\ 
  $T_p$ & Integer variable; it denotes the number of sub-carriers allocated to the spectrum path $p$\\
  ${GVD}_p$ & Integer variable; it denotes the differential delay caused by GVD on the spectrum path $p$\\
  \hline 
\end{tabular}
\caption{Variables}\vspace{-0.3cm}
\label{tab:vari}
\end{table} 
\subsection{ILP Optimization Model}
We define the objective of  the ILP model  as minimizing the usage of  sub-carriers for a spectrum allocation request, i.e.,    
\begin{equation}\label{Obj}
Minimize   \sum_{ p\in \mathcal{P}, f_i \in F, e\in E}  x_{p,e, i}  
\end{equation}
The proposed ILP model relies on the variables summarized in  Table \ref{tab:vari} and subject to the constraints defined as follows:\\ 
\noindent \textbf{\emph{Routing constraints}:}  
      Eq.\eqref{routing1} ensures that traffic on the routed path can be added and dropped only at source and destination nodes, respectively. Constraint defined in Eq.\eqref{routing2} guarantees that  a spectrum  path starts from the source node and ends at the destination node. Finally,  Eq.\eqref{routing3} eliminates the loops at source and destination nodes.\\ 
\begin{equation}\label{routing1} 
\forall p \in \mathcal{P}, \tilde{v}, v \in V, v\neq s,d :  \sum_{e=(\tilde{v},v)\in E}x_{p,e} = \sum_{e=(v,\tilde{v})\in E}x_{p,e}
 \end{equation}
\begin{flalign}
\label{routing2} &\forall p \in \mathcal{P}, \tilde{v} \in V, \tilde{v} \neq s, d:  \sum_{e=(\tilde{v}, d)\in E}x_{p,e} = \sum_{e=(s, \tilde{v})\in E}x_{p,e} = x_p  \\
\label{routing3}  &\forall p \in \mathcal{P}, \tilde{v} \in V, \tilde{v} \neq s, d:  \sum_{e=(\tilde{v}, s)\in E}x_{p,e} = \sum_{e=(d,\tilde{v})} x_{p,e} = 0   \\
\end{flalign}
%
%
\textbf{\emph{Spectrum continuity constraint}:}  In this paper, we restrict that all spectrum paths to be all-optical 
between source and destination nodes~\cite{Wang:infocom:2011}, i.e., restricted by spectrum continuity constraint.   
Eq.\eqref{continuity1} indicates that sub-carrier with index $i$ is assigned to the spectrum path $p$ from the source node.  Eq.\eqref{continuity2}   specifies that  a spectrum path can only  use sub-carriers with same index on  all fiber links it traverses. 
\begin{equation}\label{continuity1} \forall p\in \mathcal{P}, \tilde{v} \in V, \tilde{v} \neq s:  y_{p,i} = \sum_{e=(s,\tilde{v}) \in E} x_{p,e,i} 
\end{equation}
\noindent $\forall f_i \in F,  p\in \mathcal{P}, \tilde{v}, v \in V \setminus \{s,d\}:$ \\
\begin{equation}\label{continuity2}
\sum_{e=(\tilde{v},v)\in E}x_{p,e,i} = \sum_{e=(v,\tilde{v})\in E}x_{p,e,i}
\end{equation} 
\textbf{\emph{Spectrum consecutive constraints}:} For efficient modulation,   consecutive sub-carriers are required in a spectrum band when it is assigned to a spectrum path~\cite{Jinno:2009}. The spectrum consecutive constraints  are defined in Eq.\ref{reserved} and Eq.\ref{consecutive}.  Eq.\eqref{reserved} determines the number of sub-carriers allocated to path $p$. 
When two sub-carriers with index $f_i$ and $f_j$ $( j \geq  i)$ are used for $p$, the right-hand side of  Eq.\eqref{consecutive} equals to $T_p$. This constraint ensures that the gap between two sub-carriers should be equal to   or less than $T_p $.  
 When $f_i$ and $f_j$ are not used at the same time,  the right-hand side of Eq.\eqref{consecutive} results in an infinite value, which keeps Eq.\eqref{consecutive} true.\\
\noindent $\forall p\in \mathcal{P}, e\in E, v\in V, f_i \in F: $ \\
\begin{equation}\label{reserved}
 T_p= \sum_{e= (s,v)} \sum_i  x_{p,e,i}
\end{equation}
\noindent $\forall f_i, f_j \in F, j \geq i, p\in \mathcal{P}, e\in E:$
\begin{equation}\label{consecutive}
  f_j \cdot x_{p,e,j} - f_i \cdot x_{p,e,i} +1 \leq  Tp  +(2- x_{p,e,i}- x_{p,e,j}) \cdot \infty
\end{equation}
\textbf{\emph{Non-overlapping constraints}:} To avoid collision, a  sub-carrier can not be assigned to multiple spectrum paths at the same time. The binary variable $o_{p,p'}$ is defined to denote if two spectrum paths $p$ and $p'$ have at least one common link.   
    The value of  $o_{p,p'}$  is determined by Eq.\eqref{spectrum2} and Eq.\eqref{spectrum5}. When path $p$ and $p'$ share at least one fiber link, $o_{p,p'}$ equals to 1, otherwise,    $o_{p,p'}$ equals to 0.  
  Eq.\eqref{spectrum3} specifies that a spectrum slot $f_i$ can not be assigned to  $p$ and $p'$ at the same time if two paths have common links, i.e., either $y_{p,i}$ or $y_{p',i}$ can be equal to 1 when $o_{p,p'}=1$. Finally, Eq.\eqref{spectrum4} defines that spectrum assignment only happens when a path $p$ is used by the connection demand.
\begin{flalign}
\label{spectrum2} &\forall p, p' \in \mathcal{P}, p \neq  p', e \in E:   x_{p,e} + x_{p',e} - o_{p, p'} \leq 1\\
\label{spectrum5} &\forall p,p' \in \mathcal{P}, e\in E:           o_{p,p'} \leq \sum_e   x_{p,e}\cdot x_{p',e}\\
\label{spectrum3}  & \forall p,p' \in \mathcal{P}, p \neq p', f_i \in F:  y_{p,i} + y_{p',i} + o_{p,p'} \leq 2\\
\label{spectrum4} & \forall p \in P, f_i \in F: x_{p} - y_{p,i} \geq 0
\end{flalign} 
The constraint defined in Eq.\eqref{spectrum5} is non-linear. We define a  new binary variable denoted as $\gamma_{p,p',e} =x_{p,e}\cdot x_{p',e}$, hence Eq.\eqref{spectrum5} is linearized as follows:
\begin{flalign}
\label{line0} & \forall p,p' \in \mathcal{P}, e\in E:   o_{p,p'} \leq \sum_e  \gamma_{p,p',e}\\
\label{line1} &  \forall p,p' \in \mathcal{P}, e\in E:   \gamma_{p,p',e} \leq x_{p,e}\\
\label{line2} &  \forall p,p' \in \mathcal{P}, e\in E:   \gamma_{p,p',e} \leq x_{p',e}\\
\label{line3} & \forall p,p' \in \mathcal{P}, e\in E:    x_{p',e} +x_{p,e}- \gamma_{p,p',e} \leq 1
\end{flalign}
\textbf{\emph{Guard-band constraint}:} 
 The constraint defined in Eq.\eqref{guard-1} specifies that the spectrum assignment only happens when the available sub-carriers are sufficient to meet the guard-band requirement.   When a sub-carrier $f_i$ is allocated to a path $p$, a sub-carrier within the range $\{f_i-GB, f_i+GB\}$ cannot be allocated to other paths, i.e., all sub-carriers within the range  $\{f_i-GB, f_i+GB\}$ are excluded from the available spectrum set of link $e$ for other paths.
Eq.\eqref{guard-2} ensures that a guard-band exists between  two spectrum paths $p$ and $p'$,  if they share at least one common link.  When $p$ and $p'$ have no 
common link,  $o_{p,p'}$ equals to 0, which guarantees that  Eq.\eqref{guard-2} is always true. \\
  \begin{equation}\label{guard-1} 
  \forall p \in \mathcal{P}, e \in E, \{f_i \pm GB\} \in F \setminus F^e:  x_{p,e,i} =0
 \end{equation}
 \noindent $\forall p, p' \in \mathcal{P}, e\in E,   f_i, f_j \in F:$\\
 \begin{equation}\label{guard-2}  
  |f_j \cdot x_{p,e,j} - f_i \cdot x_{p',e,i}| \geq GB \cdot o_{p,p'} 
\end{equation}\\
\textbf{\emph{ Bandwidth constraint}:} This constraint ensures that the number of  sub-carriers  assigned to all the spectrum paths for connection demand $R$ is equal to the traffic demand $T_r$, 
\begin{equation}\label{traffic}
 \sum_{ p\in \mathcal{P}, f_i \in F}  y_{p,i} = T_r
\end{equation}
\textbf{\emph{Differential delay constraint}:} 
Differential delay in OFDM-based networks are mainly caused by the fiber effects and path diversity, as it is discussed in Sec.II B. We denote the  maximum differential delay caused by the GVD on path $p$  as $GVD_p$. $GVD_p$ is calculated based on the Eq.\eqref{GVD}. As shown in Eq.\eqref{GVD5}, $T_p \cdot s_f$ is the gap between the highest frequency and the lowest frequency of a spectrum band. $\sum_e L_e \cdot x_{p,e}$ calculates the length of the path $p$. 
 \begin{equation}\label{GVD5}
  \forall p \in \mathcal{P}, e\in E:     {GVD}_p = D(f_c)\cdot s_f \cdot T_p \cdot \sum_e L_e \cdot x_{p,e}
  \end{equation}
Note that Eq.\eqref{GVD5} is non-linear. We hence define an integer variable, denoted as $z_{p,e}$, $z_{p,e}=T_p \cdot x_{p,e}$. The GVD constraint can be linearized as shown in Eq.\eqref{GVD5line}.
 \begin{equation}\label{GVD5line}
  \forall p \in \mathcal{P}, e\in E:     {GVD}_p = D(f_c)\cdot s_f \cdot \sum_e L_e \cdot  z_{p,e}   
  \end{equation}
The variable $z_{p,e}$ is restricted by the constraints defined in  Eq.\eqref{GVD2} Eq. \eqref{GVD3} and 
Eq.\eqref{GVD4}, which determine that the value of $z_{p,e}$ is either zero or equal to $T_p$.   The delay of a path $p$ is defined in Eq.\eqref{delay}. Hence the total differential delay is defined in Eq.\eqref{DD}, which can not   exceed the buffering capacity of the electronic layers.  i.e., the differential delay between any two spectrum paths used for a connection can not exceed $M$ (Eq.\eqref{DD}).
\begin{equation}\label{GVD2}
\forall p \in \mathcal{P}, e\in E:   z_{p,e} \leq x_{p,e} \cdot |F|
\end{equation}
\begin{equation}\label{GVD3}
\forall p \in \mathcal{P}, e\in E:   z_{p,e} \leq T_p
\end{equation}
\begin{equation}\label{GVD4}
\forall p \in \mathcal{P}, e\in E:   z_{p,e} \geq T_p -(1-x_{p,e})\cdot |F|
\end{equation}
 \begin{flalign}\label{delay}
  pd_p = \sum_{e\in p} LD_e 
\end{flalign} 
 \begin{flalign}\label{DD}
 \forall p, p' \in \mathcal{P}:  |pd_p - pd_{p'}|+(GVD_p + GVD_{p'}) \leq M 
\end{flalign}
\textbf{\emph{Discussion:}} 
\subsubsection{Problem Size}
The complexity of an ILP formulation is known to be exponential, i.e. $O(2^n)$, where $n$ is the number of variables. Thus the
proposed ILP model for  multiple path computation and spectrum assignment for parallel transmission  has an exponential complexity with $n$ in $O(|P|\cdot(|P|+|E|\cdot|F|))$, where $|P|$ is the number of used paths, $|E|$ and $|F|$ are number of links and number of sub-carriers respectively. It 
makes the optimization model computationally  expensive and rather infeasible in practice.  
Take a simple 
example, for a network of size $|V|=15$, $|F|=16$, there are $|V|\cdot |V-1|= 210$ node pairs. For each node pair, there are $|P|\cdot |F|$ instances of $y_{p,i}$.  
Assume  four paths are used in the connection, i.e., $|P|=4$, we have 13440 $y_{p,i}$ variables.  Other variables can be calculated in a similar way. 
Thus, the total number of variables in  the ILP model is very high even for small networks. 
\subsubsection{Complexity Reduction}
The problem size can be reduced by pruning the variables.  A common method used in the literature is to compute a set of paths  in advance and use them as input to the ILP model.
 However, the solutions are limited in the pre-computed path set in this case and the  complexity of the advance path computation should also be considered.

\subsection{Heuristic Algorithm} 
 In this section, we propose a heuristic algorithm which decomposes the parallel transmission problem into two sub-problems, i.e.,  multipath computation and spectrum assignment,  as shown in Alg.\ref{alg:RSA}.
 The objective and constraints of  ILP model are also considered in the   heuristic algorithm.  In addition, we limit the maximum number of fiber-level paths in order to reduce the complexity, i.e., maximum $K$ fiber-level paths are computed for a given traffic demand.  

\begin{algorithm}
\small
      \KwIn{$G(V, E), K$, $R(S,D,T_r)$}
   \KwOut{One or multiple spectrum paths for $R$}
   \textbf{Phase 1: Multiple Fiber-level Path Computation}  \\
 \While{($|\mathcal{FP}| \leq K$)}{  
 \While {$destination(fp)$$\neq D$}  {
     Select min-delay path $fp$ from $\mathcal{S}$ \\
  \For {all nodes $v'$ connected to $destination(fp)$}{
       \If {($v'$ not traversed in $fp$)}{
           create $fp'$ by extending $p$ to $v'$ \\
               add $fp'$ to $\mathcal{S}$
      }
    Put $fp$ into $\mathcal{FP}$\\
  Remove $fp$ from $\mathcal{S}$ 
  }
}       
}
Return $\mathcal{FP}$\\
 \textbf{Phase 2: Spectrum Assignment}  \\
//Step 1: Single spectrum path first;\\
 	   	  \For{$k =1$  to $K$, $fp_k \in \mathcal{FP}$}{
 	   	  Identify the spectrum path with maximum consecutive sub-carriers, i.e., $p_k$;\\
 	   	  \If{$F(p_k) \geq T_r $}{
 	   	  A single spectrum path found;
 	   	  break;
 	   	  }
}
//Step2: Multiple Spectrum Paths\\	  
\For{$k =1$  to $K$, $fp_k \in \mathcal{FP}$}{
\For {all $e_i \in fp_k$}{
 	  	Find spectrum paths on the fiber-level path $fp_k$ and put in the path set $\mathcal{P}_k$
 	  	} 
 \For {$k=1$ to $K$, $fp_k \in \mathcal{FP}$}{
 Sort all available spectrum paths  in the increasing order of delay; and put in path set $\mathcal{P}$\\
  $N= |\mathcal{P}|$\\
 }
 \For{$k= 1$ to  $N$} {
    \If{$ pd_{p_{k}}- pd_{p_1} \leq M$}
  {
    $F+= F_{p_k}$\\
       \If {$F \geq T_r$}
  {
  Return spectrum paths and  break;
 }
 }
 }
 }
  \caption{RSA for Parallel Transmission} 
 \label{alg:RSA}
\end{algorithm}  

\subsubsection{Multipath Computation}
The first phase of the proposed heuristic algorithm is to compute a set of fiber-level paths which are used as input to the spectrum 
assignment (phase 2 in Alg.\ref{alg:RSA}).  The algorithm starts from collecting all paths originating from source node $S$.  
All outgoing links from $S$ are placed in a set denoted as $\mathcal{S}$ and sorted in an increasing order of path delay. The shortest 
path in $\mathcal{S}$, denoted as $fp$, is selected and extended to all the nodes connected to the sink node of $fp$, i.e., $destination(fp)$. 
Afterwards, the path set $\mathcal{S}$
is updated with the extended links and  the shortest path from current $\mathcal{S}$ is selected. The same procedure is repeated 
till the shortest path in $\mathcal{S}$ reaches the destination node $D$. The computed path $fp$ is placed in fiber-level path set 
$\mathcal{FP}$ and removed from $\mathcal{S}$.
The algorithm continues to select the shortest path from the updated $\mathcal{S}$ and repeats the path computation. It breaks when no path can be computed or $K$ fiber-level paths have been computed.  
In the worst case scenario, the phase 1 of Alg.\ref{alg:RSA} has to visit all the nodes in the network to find a path $fp$ between $S$ and $D$. Assume the maximum node degree in the network  is $Deg(V)\}$, the complexity of multipath computation phase of  Alg.\ref{alg:RSA} is in $O(|V|^2 \cdot Deg(V) \cdot K) $.
 
\subsubsection{Spectrum Assignment}
In phase 2 of Alg.\ref{alg:RSA}, the computed path set, i.e., $\mathcal{FP}$, is used as input. The algorithm tries to find a single path solution first.  It identifies the spectrum path with maximum number of consecutive sub-carriers on each fiber-level path $fp\in\mathcal{FP}$ and compares the available bandwidth with $T_r$.  
When it fails to find a single path  solution, the algorithm continues to find a multipath solution, i.e., aggregating 
spectrum fragments from multiple spectrum paths. All spectrum paths in $\mathcal{FP}$ are sorted in the increasing 
order of delay and put in the set $P$. Afterwards, the differential delay and bandwidth constraints are  checked. If  the 
differential delay between a spectrum path $p_k \in \mathcal{P}$ and the shortest path $p_1 \in \mathcal{P}$ is no 
larger than $M$, i.e., $pd_{p_k} -pd_{p_1} \leq M$, $p_k$  is included in the solution. Note that the GVD caused differential delay is not considered here. The algorithm outputs a solution 
when bandwidth requirement is satisfied.   In the worst case scenario, the phase 2 of Alg.\ref{alg:RSA} has to check 
all the sub-carriers over all fiber links. Hence, the computational complexity of spectrum assignment phase is in $O(|K|\cdot |F|\cdot |E|)$.  
 
\section{Performance Evaluation}\label{performance}
\par
\begin{figure}
	\centering
	\includegraphics[width=1\columnwidth]{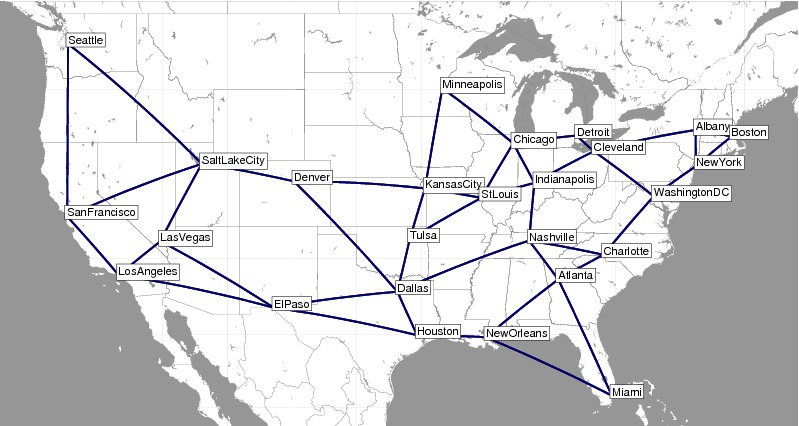} 
  \caption{US backbone network topology \cite{SNDlib} under study with 24 nodes and 84 links} \label{fig:us} 
  \vspace{-0.3cm}
 \end{figure}

 In this section, we evaluate the performance of the proposed parallel transmission algorithms, under the dynamic traffic conditions. The connection requests arrive following a Poisson process with an average arrival rate of $u$ and 
the holding time of each connection request follows the negative exponential distribution with an average value of $h$ 
time units; thus the traffic load in the network is quantified as $u*h$ in $Erlang$.  
In our study, the mean inter-arrival time of the connection requests is $1$ time unit and mean holding time is varying 
to achieve different network loads.  The measure of blocking probability, as a common metric for assessing the performance, is defined as the percentage of blocked connection requests versus total connection requests. The source and destination pair is randomly selected in the network 
under study.    Fig.\ref{fig:us} 
shows the network topology used in this paper \cite{SNDlib}.

\par The  ILP model  is implemented in Gurobi Optimizer \cite{gurobi} and the heuristic algorithm 
is implemented with a Java event-driven simulator.  
  We consider two representative values as the maximum acceptable differential delay, i.e., 
 $250\mu s$ as suggested in ITU-T G.709 \cite{ITU-T:G.709}  and $128ms$ which can be supported by commercial framer 
 mappers\cite{Intel:datasheet}.  The main results shown in this section are evaluated on the US backbone topology as shown in Fig.\ref{fig:us} and  have a confidence interval of 95$\%$.      
      We study the impact of different number of sub-carriers per fiber ($|F|$) and  the maximum number of fiber-level paths ($K$) as well as maximum acceptable differential delay ($M$) on the proposed algorithms.

\subsection{ILP Model Evaluation} \label{small}
 \par Given the complexity issue of ILP optimization, we first evaluate the proposed ILP model in a scaled-down network scenario where an optimal solution can be obtained within a reasonable time. The number of sub-carriers per fiber link is 16; the maximum differential delay  is  $128ms$ and guard band is set to be one sub-carrier.
The heuristic algorithm is also evaluated in the  same experimental setting and compared with the ILP model.  
Afterwards,  we study only the heuristic algorithm in a scenario where the ILP model becomes intractable and thus of little practical relevance.     The traffic load is generated by the connection requests following Poisson process with bandwidth requirement uniformly distributed between 1   and 4 sub-carriers. When the  network is stable at a certain network load, a  connection demand requesting between 4 and 6  sub-carriers is sent to a randomly selected source and destination pair and an algorithm is invoked. The same experiment is repeated over 50 times to obtain a mean value.
  
\par    Table \ref{tab:block-ilp} shows the percentage of blocked connections  at each given network load. To study the impact of pre-computed path set,   the maximum number of fiber-level paths that can be used in the parallel transmission in the heuristic is set to be 10 and 40, respectively. There is no limitation of number of fiber-level paths used in the ILP evaluation, it stops when the model either finds a solution, or it is time out.   It can be seen that ILP model always outperforms the heuristic algorithm when the problem is tractable. For instance, none of  the connection requests is blocked using ILP model  when network load is $30 Erlang$. However,    $16.0132\%$ and $8.0723\%$  connections are blocked using the heuristic algorithm with $K$ =10 and  $K$ = 40, respectively. When the network load increases,  the number of blocked connections increases with both the ILP model and the  
    heuristic.  However,   the performance of proposed heuristic algorithm is getting close to the ILP model when the 
    pre-computed paths are sufficient. For instance, mutipath routing with a set of 40 precomputed fiber-level paths leads to      around $22.753 \%$ blocking at $45 Erlang$ while the ILP model leads to $20.902 \%$ blocking at the same     network load. 
    
\par This set of results shows that the main disadvantage of the proposed heuristic is the issue of pre-computed paths set. However, the heuristics can always find a solution in a reasonable time, which may be single or multiple spectrum paths. At the same time,  we have observed that the ILP becomes infeasible  when high network load or with more sub-carriers per fiber link in the network.
     
\begin{table}
 \centering
   \caption{Connection blocking probability with the ILP model and Heuristic in US network } \label{tab:block-ilp}
        \begin{tabular}{ | c | c | c | c |}
        \hline
         Load   & Heuristic &   Heuristic   & ILP \\
         ($Erlang$) & ($K$= 10) & ($K$= 40)  & $M=128ms$ \\ \hline
  30&        16.0132\% &	8.0723\% &	0.000\% \\ \hline
35&	24.823\%	&10.419\%	&6.025\%\\ \hline
	40&34.392\%&	18.362\%	&14.025\%\\ \hline
	45& 40.015\%&	22.753\%	&20.902\%\\ \hline              
        \end{tabular}
\end{table}

 \subsection{Performance of the Heuristic Algorithm} \label{large}
 In this section, we investigate the performance of  proposed heuristic algorithm. The simulation 
 parameters are summarized in Table \ref{tab:para}. Connection requests are uniformly distributed on the randomly selected source and destination nodes. Average 20,000 connection requests are generated at every network load; and  every experiment is repeated  5 times to obtain an average value.   We distinguish  two parallel transmission scenarios which are defined as follows:
 \begin{definition}
 \emph{Parallel transmission on single spectrum path (ST)}: The required sub-carriers are allocated on the  shortest spectrum path.  Differential delay issue is hence not considered.    
 \end{definition}
 \begin{definition}
 \emph{Parallel transmission on multiple spectrum paths (PT)}: Here, the required sub-carriers are not restricted to a single spectrum path. In general, multiple spectrum bands from a single fiber, or from multiple fiber links can be used. In the results shown in this section, the heuristic algorithm with maximum allowable differential delay $128ms$ and $250\mu s$ are denoted as PT-1 and PT-2, respectively. 
 \end{definition}
 
 \begin{table}
 \centering
   \caption{Simulation Parameters } \label{tab:para}
        \begin{tabular}{ | l | l | }
        \hline
         F: number of frequency slots per link &128 \\
        \hline
  GB: number of sub-carries  as guard-band&  0-3\\ \hline
 $Tr$: number of requested sub-carriers& 5,10,15\\ \hline
 $K$: maximum number of fiber-level paths &30  \\ \hline
PT-1:   with large buffer & $M=128ms$\\
PT-2:   in line with ITU-T G.709& $M=256\mu s$	 \\ \hline              
        \end{tabular}
\end{table}

\subsubsection{Blocking Probability  vs. spectrum requirements and network loads} 
\begin{figure}
        \centering
        \begin{subfigure}[b]{0.47\textwidth}
                \centering
                \includegraphics[width=\textwidth]{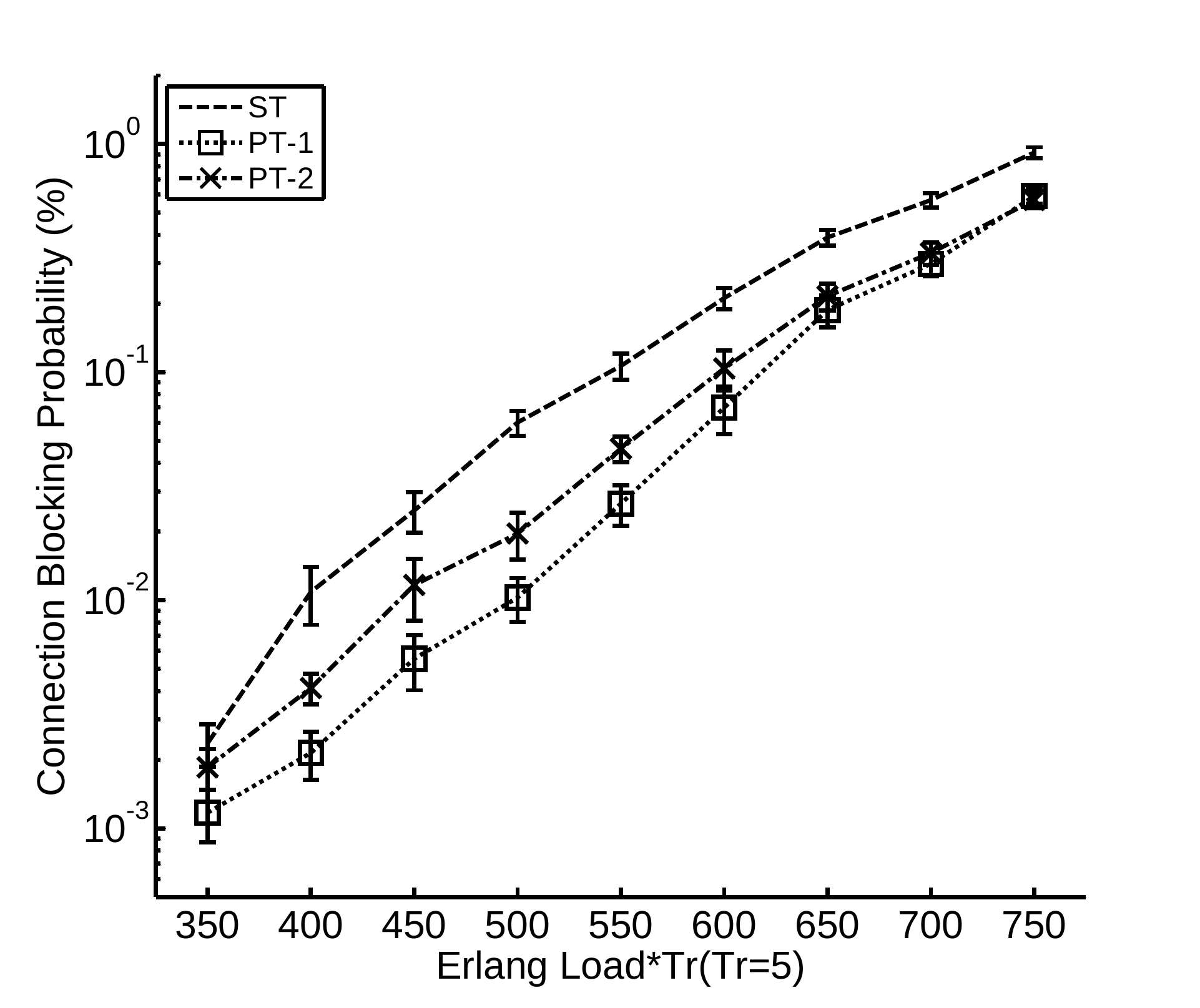}
                \caption{ GB=0, Tr=5 }
                \label{fig:gb0-1}
        \end{subfigure} \\
        ~ 
        \begin{subfigure}[b]{0.47\textwidth}
                \centering
                \includegraphics[width=\textwidth]{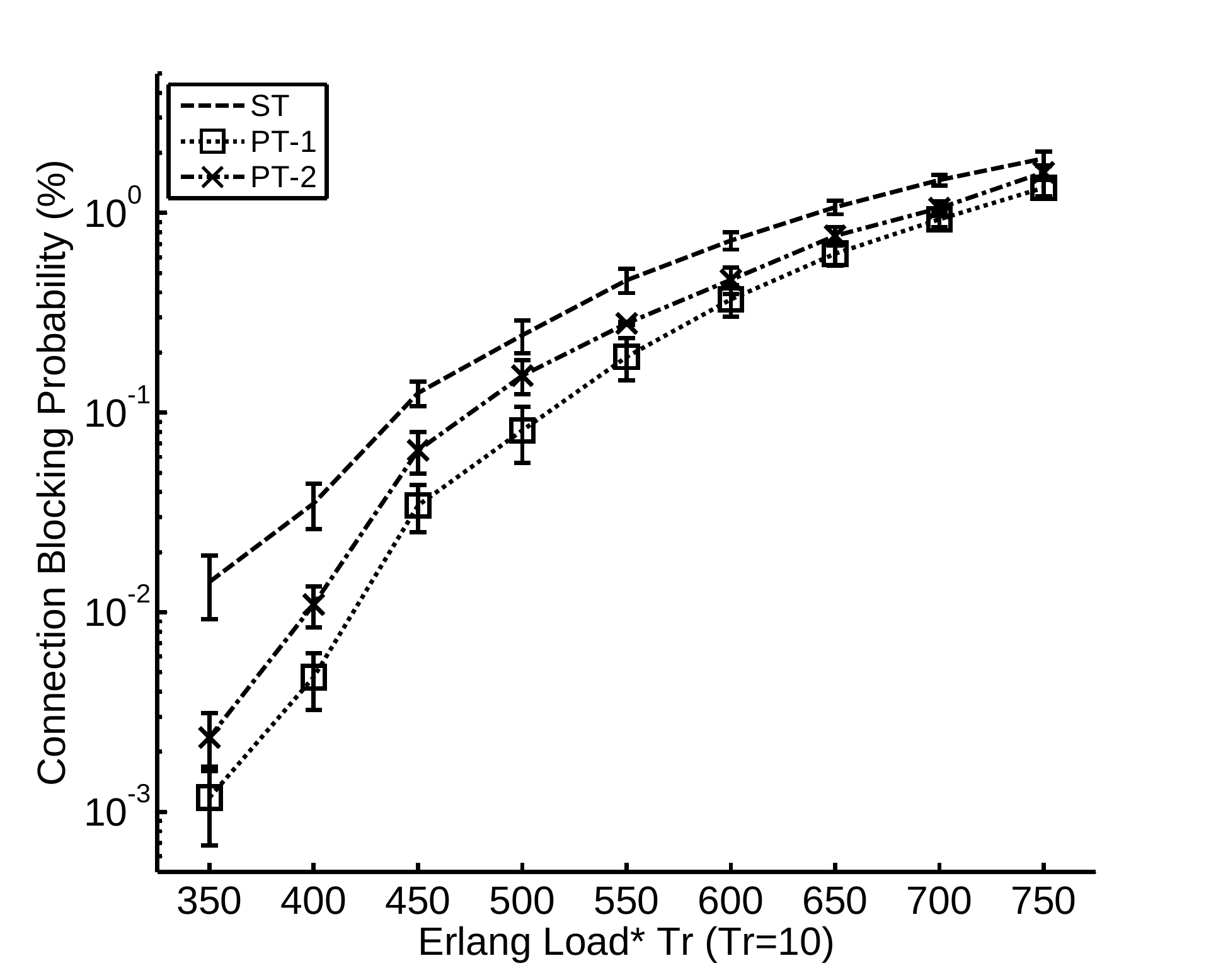}
                \caption{ GB=0, Tr=10}
                \label{fig:gb0-2}
        \end{subfigure}\\
        ~ 
        \begin{subfigure}[b]{0.47\textwidth}
                \centering
                \includegraphics[width=\textwidth]{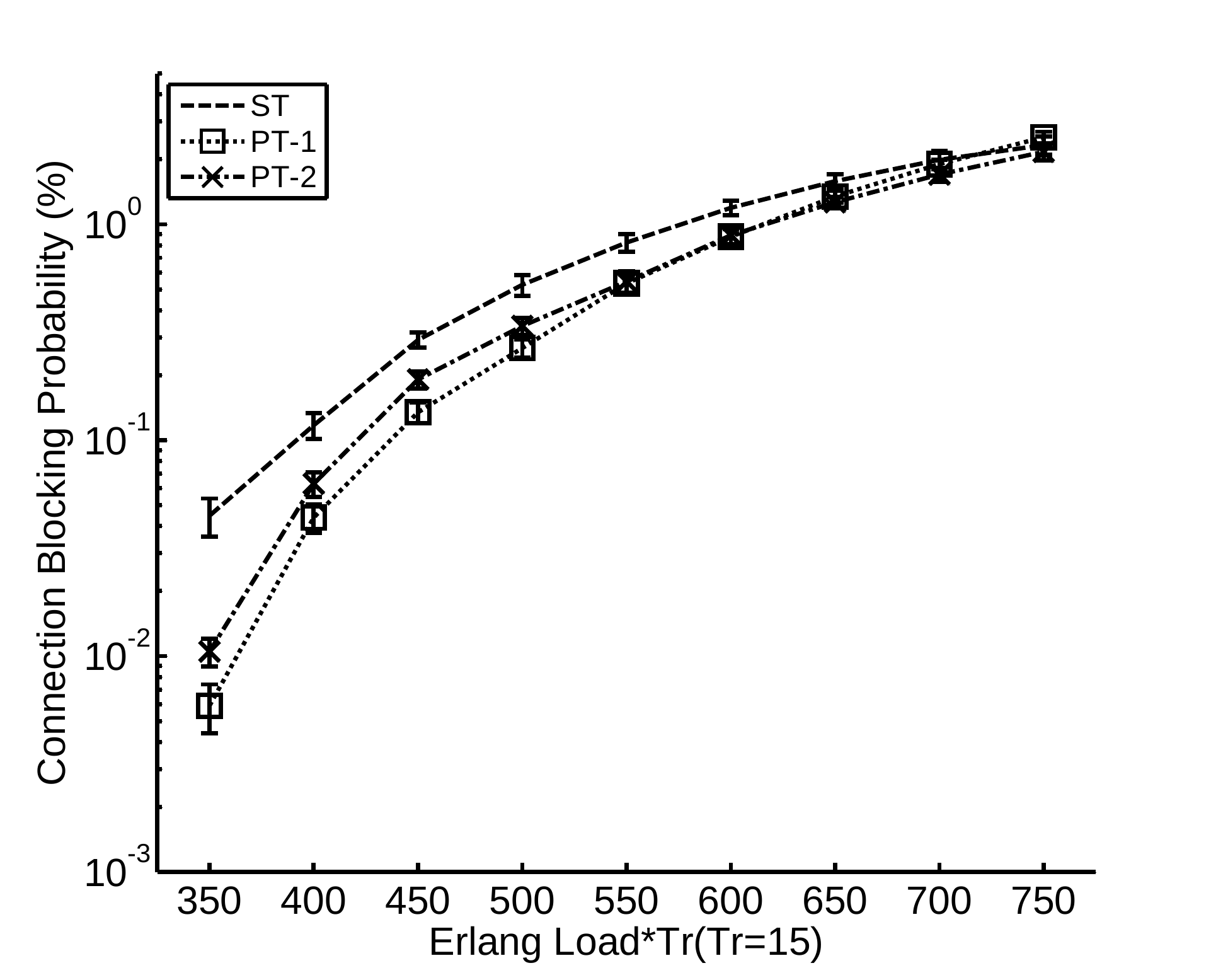}
                \caption{GB=0, Tr=15}
                \label{fig:gb0-3}
        \end{subfigure}
        \caption{Blocking probability vs. network load, with no guard-band (GB=0)}\label{fig:gb0} 
\end{figure}

Fig.\ref{fig:gb0} shows the 
blocking probability of parallel transmission with different spectrum requirements. Here, we assume there is no guard-band between adjacent spectrum paths, i.e., $GB=0$. It can be seen that allowing for multiple spectrum paths in the 
parallel transmission can increase the acceptance ratio of connection requests, leading to the reduction of blocking 
probability.  With the increase of the spectrum requirement, the blocking probability also increases. For instance, the 
$1.8\%$ connection requests are blocked with $T_r=10$ restricted to a single spectrum path (ST) at traffic load of $75 Erlang$. With $T_r=5$, maximum $0.9\%$ connection requests with SP at traffic load of $150 Erlang$. Regardless of the spectrum requirements and network loads, parallel transmission over multiple spectrum paths, i.e., PT-1 and PT-2, can always lead to a lower blocking probability, due to the fact that spectrum fragments are now aggregated instead of ``wasted" in case of ST. When network load is very high, the performance of PT is also getting worse, especially when connection granularities are large. It will eventually result in the same performance as ST, as shown in Figs. \ref{fig:gb0-2} and \ref{fig:gb0-3}.

\subsubsection{Impact of Differential Delay Constraint}
The compensation of differential delay in the upper layers can significantly  affect the performance of parallel 
transmission in the underlying optical network. We define PT-1 and PT-2 with maximum acceptable differential 
delay as $128ms$ and $250\mu s$, respectively. As it is discussed in Sec.II-A, for a connection with length of $1\times 
10^4$km, the maximum differential delay in a standard single mode fiber is only $50ms$. Hence,  the differential delay 
constraint of PT-1 is sufficient to cross the longest path in the studied network. It leads to the fact that PT-1 can utilize 
longer paths comparing with PT-2, thus leading to the lower blocking probability, as shown in Fig.\ref{fig:gb0}. With the 
same experimental settings, e.g., $T_r=10$ and $75 Erlang$, blocking probability of PT-1 is $0.2\%$ less comparing 
with PT-2. However, since PT-1 consumes more spectrum resources by using longer paths, and it will fail to find a solution 
for a request requiring large number of sub-carriers when network load is high. As shown in Fig.\ref{fig:gb0-3}, the 
blocking probability of PT-1 is $0.4\%$ higher than PT-2 at network load of $50Erlang$.

\subsubsection{Impact of Guard-band}
\begin{figure}
        \centering
        \begin{subfigure}[b]{0.47\textwidth}
                \centering
                \includegraphics[width=\textwidth]{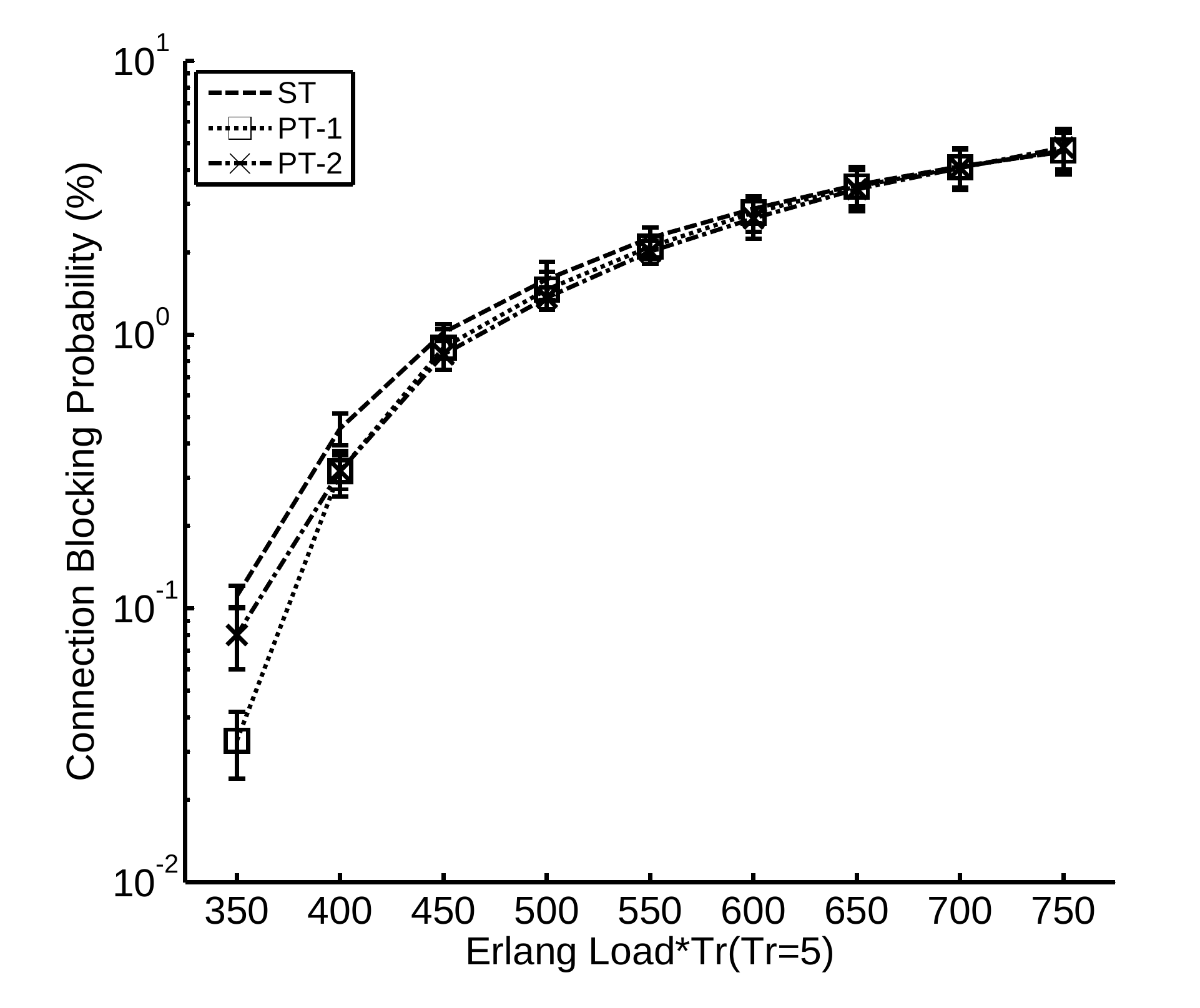}
                \caption{ GB=3, Tr=5 }
                \label{fig:gb3-1}
        \end{subfigure}  \\
                \begin{subfigure}[b]{0.47\textwidth}
                \centering
                \includegraphics[width=\textwidth]{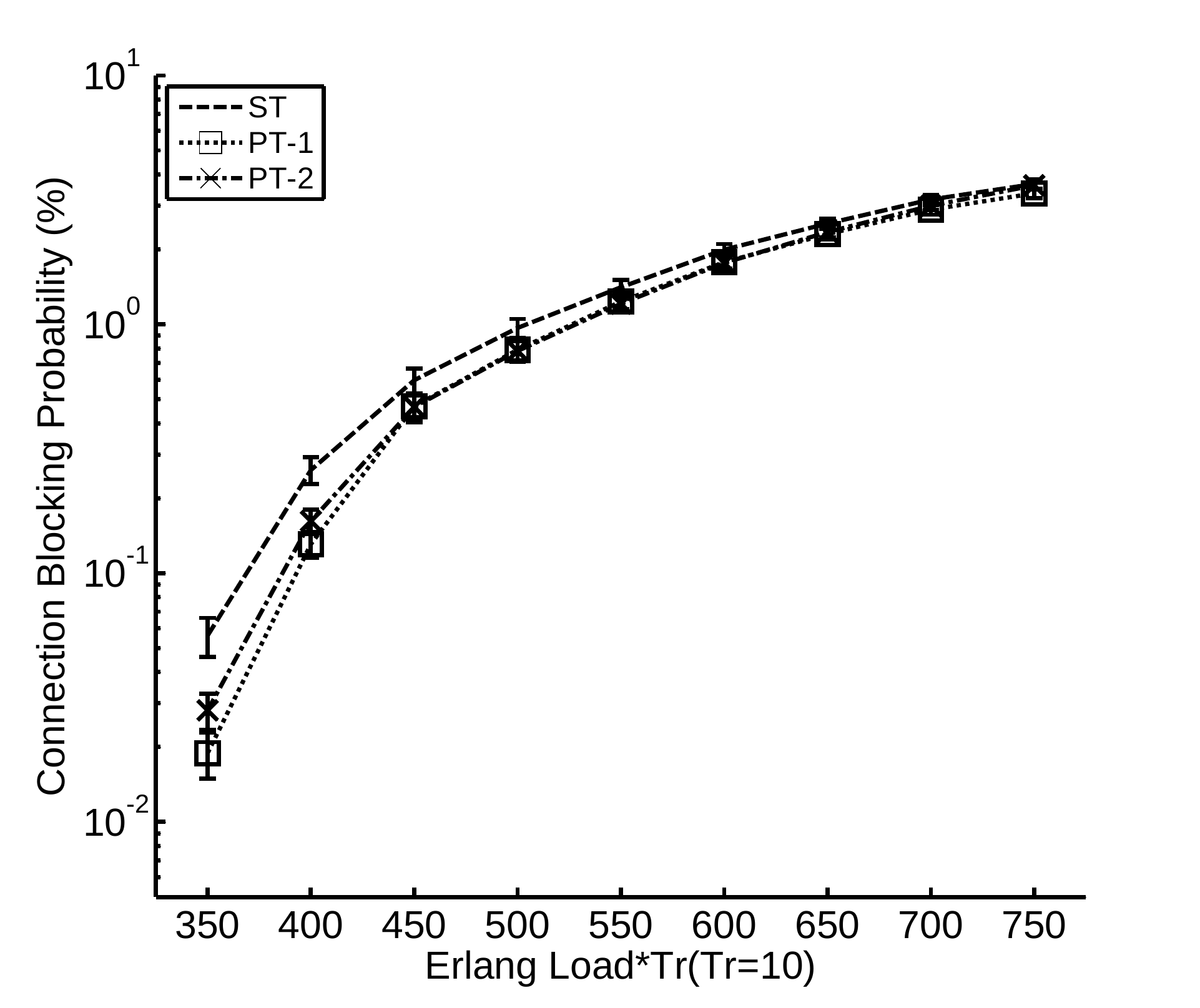}
                \caption{ GB=3, Tr=10}
                \label{fig:gb3-2}
        \end{subfigure}  \\
                \begin{subfigure}[b]{0.47\textwidth}
                \centering
                \includegraphics[width=\textwidth]{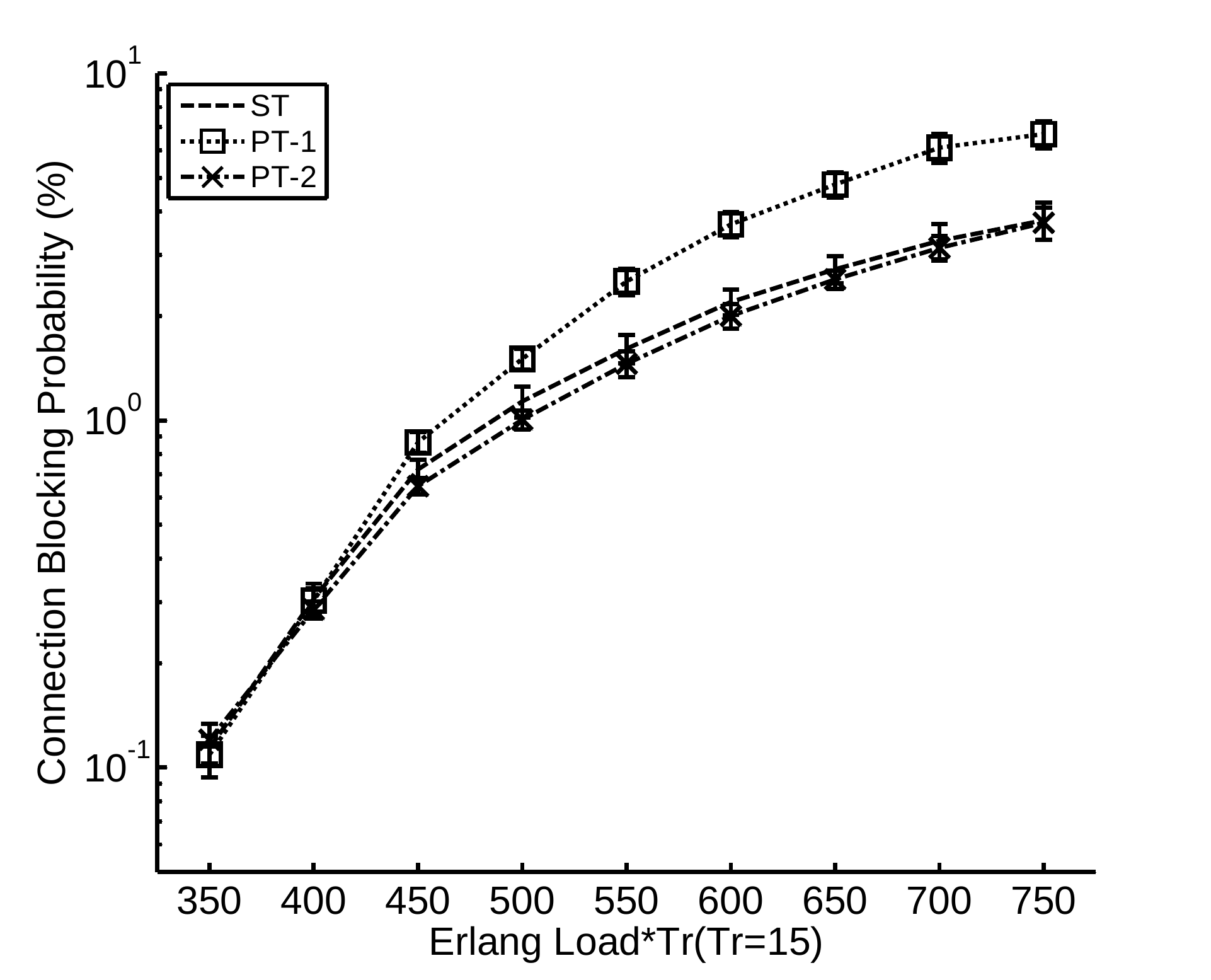}
                \caption{GB=3, Tr=15}
                \label{fig:gb3-3}
        \end{subfigure}
        \caption{Blocking probability vs. network load, with a large guard-band (GB=3)}\label{fig:gb3}
\end{figure}

When the guard-band is required, parallel transmission with multiple spectrum paths 
may consume more resources on guard-bands, counteracting its benefits. Fig.\ref{fig:gb3} shows the performance of 
parallel transmission with single and multiple spectrum paths in terms of blocking probability.  It can be seen that both 
PT-1 and PT-2 can still reduce blocking probability, comparing with single spectrum path only (ST). However, with 
increasing network load, the resource consumed by guard-bands required by multiple spectrum paths  leads to the 
degradation of performance, despite the PT aggregating spectrum fragments. As shown in Fig.\ref{fig:gb3-1} and 
Fig.\ref{fig:gb3-2}, PT-1 has the best performance when network load is low. When the network load is high, e.g., 
$75Erlang$ and $150Erlang$ with $T_r=10$ and $T_r=5$, respectively, both PT-1 and PT-2 have almost the same 
blocking probability as the ST. 

When spectrum requirements are high, e.g., $T_r=15$, it is not easy to find spectrum paths for both ST and PT. 
While PT-1 tends to reserve more resources by allowing for long paths, it has the highest blocking probability with  $GB=3$, as shown in Fig.\ref{fig:gb3-3}. However, PT-2  has very strict differential delay constraint, i.e., $250\mu s$, it leads to the similar performance as SP with the increasing network load.  It is due to the fact that PT-2 can only find a single spectrum  path as the solution under this network condition. 
\subsubsection{Spectrum Fragments Aggregation Ratio}
\begin{figure}
\center {
\includegraphics [width = 0.9\columnwidth] {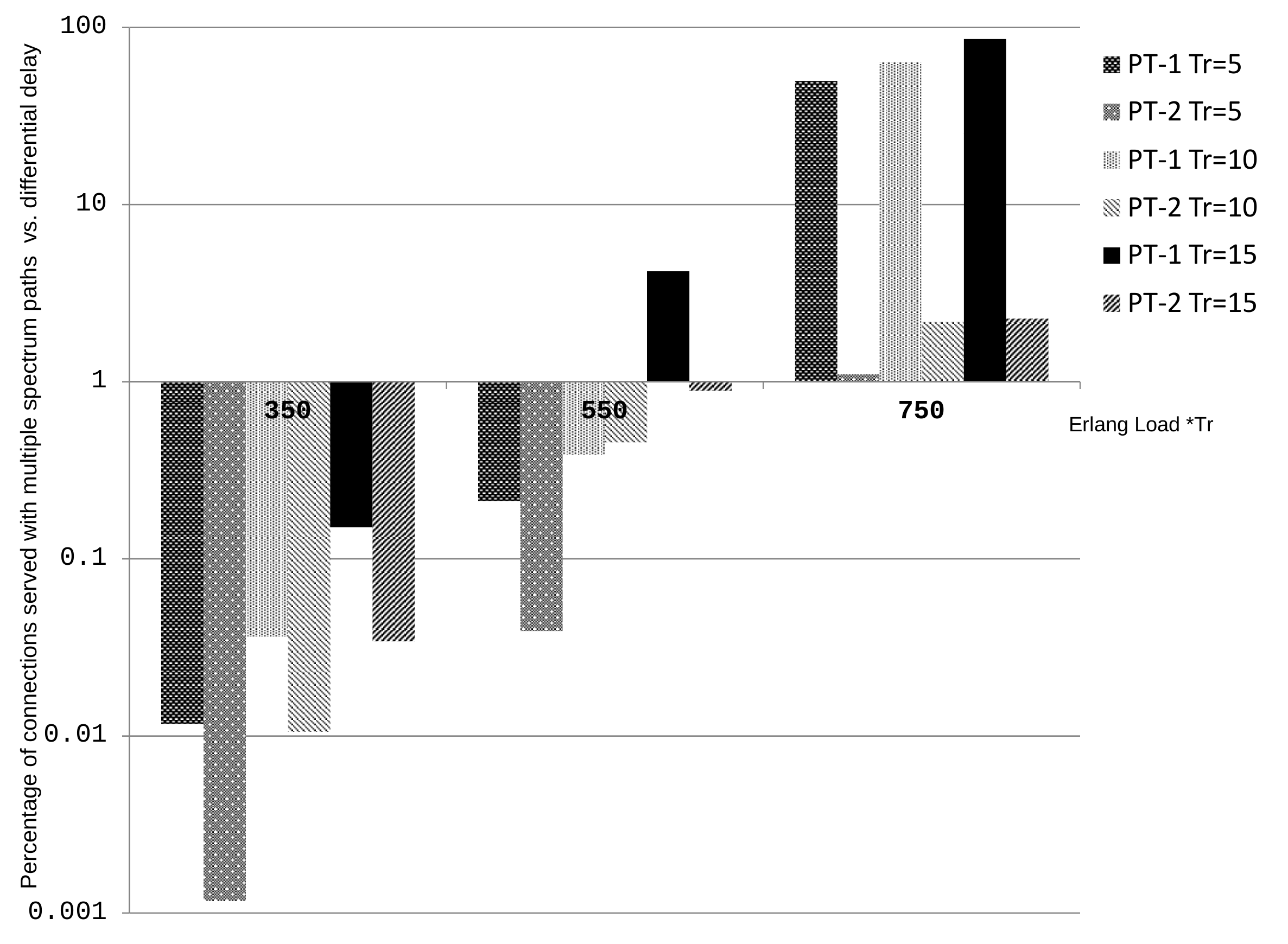}
} \caption{Percentage of connections served with multiple spectrum paths in parallel transmission vs. differential delay vs. Erlang Load * Spectrum requirement (Tr), GB=0}
\label{resfigMPPercent}\vspace{-0.3cm}
\end{figure}

As it has been discussed before, the proposed parallel transmission algorithm tries to aggregate spectrum fragments 
when it can not find a single spectrum path.  Hence, the fraction of connections that are served with multiple spectrum paths implies how often the spectrum fragments are aggregated.  
\begin{definition}
\emph{Spectrum fragments aggregation ratio} is defined as the percentage of connection requests served with multiple spectrum paths when parallel transmission algorithms are applied. It is affected by the bandwidth requirements and differential delay constraints.  
\end{definition}
Fig. \ref{resfigMPPercent} presents measurements under three different loads (Erlang load * spectrum requirement ($T_r$)). 
As it can be seen from Fig.\ref{resfigMPPercent}, the restrict differential delay constraint in PT-2 ($250\mu s$) results in less connections using multiple spectrum paths. At low and medium loads ($Erlang load*T_r=350$ and 550, respectively),  most requests ($>$ 95 \%) are served with a single spectrum path and in the case of PT-2, the remaining requests are served mostly ($>$ 98 \%) using only 2 spectrum paths. In case of medium load ($Erlang load*T_r=550$) with a large spectrum requirement($T_r=15$),  PT-1 aggregates spectrum fragments more frequently comparing with PT-2. Around $55\%$ connection requests are served with 2 spectrum paths and   $25 \%$ connections are using 3 spectrum paths.  $11 \%$ connections are set up with  4 spectrum paths and the others use   higher number of spectrum paths. In case of PT-1, maximum 10 spectrum paths were used for connection requests. 

\begin{figure}
\center {
\includegraphics [width = 0.9\columnwidth] {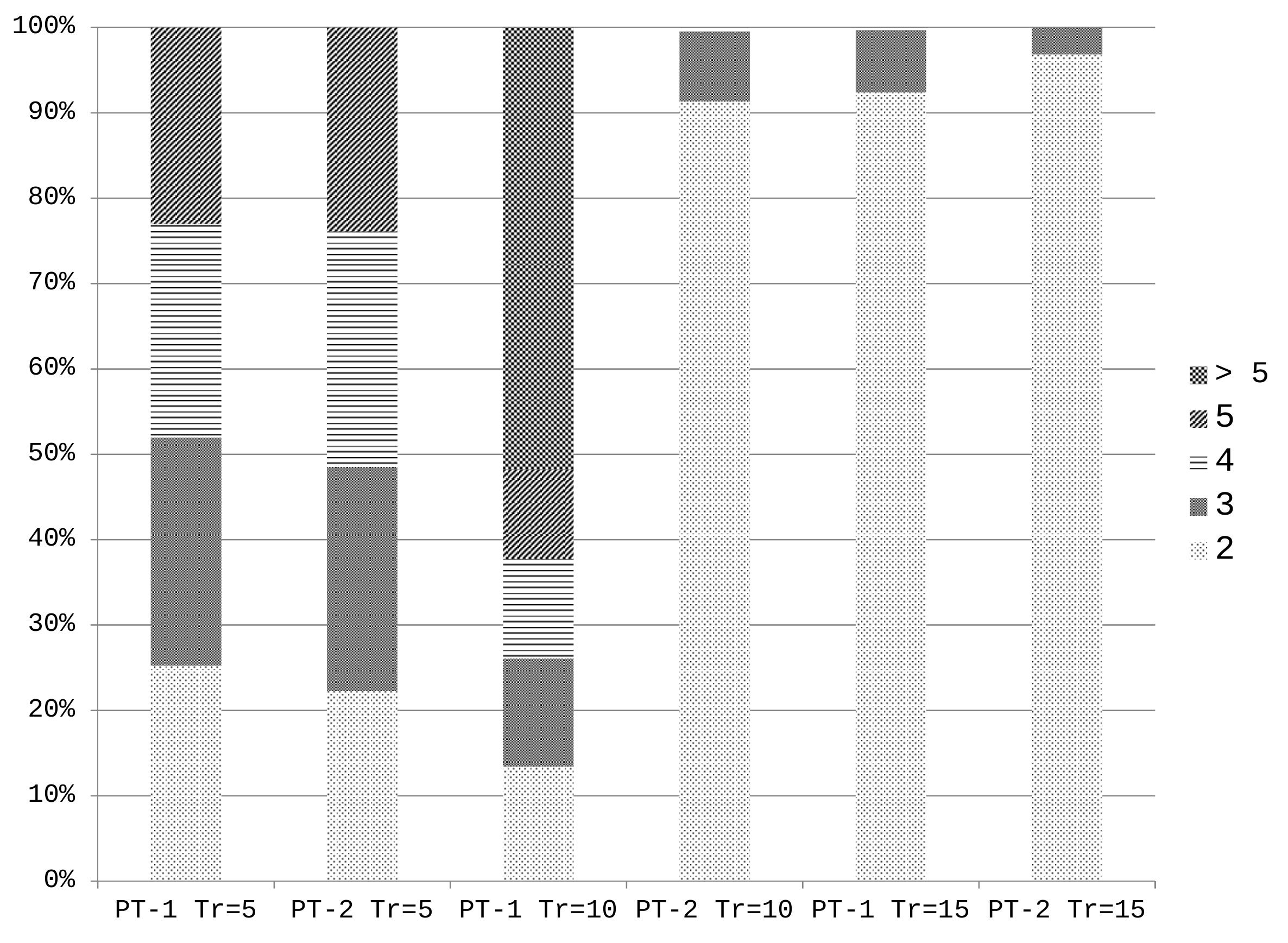}
} \caption{Percentage Distribution of Connections in parallel transmission vs. number of spectrum paths at a high load (Erlang Load *Tr=750), GB=0  }\vspace{-0.2cm}
\label{resfigSPDist}
\end{figure}
 This behavior is even more pronounced at high network loads: the distribution of the requests served by multiple 
 spectrum paths vs. the number  of spectrum paths used is presented in Fig. \ref{resfigSPDist}. As it can be seen in Fig.\ref{resfigSPDist}, PT-2  uses only 2 spectrum paths for most connection requests, while a small fraction of 
 connections using 3 spectrum paths. However in the case of PT-1, as the requirement for spectrum slots increases, the 
 fraction of connections using more spectrum paths also increases, with some connections also recorded using as many 
 as 15  spectrum paths, each with a single spectrum slot.  The unbounded nature of the number and  length of the 
 spectrum paths in PT-1 imply that  algorithm can provision spectrum slots across disproportionately longer paths as 
 compared to the single path algorithm (ST). As a result, it leads to blocking of future connection requests, thereby 
 degrading the performance of the algorithm at high network loads.

 \subsubsection{Impact of Network Topologies}
\begin{figure}
	\centering
	\includegraphics[width=1\columnwidth]{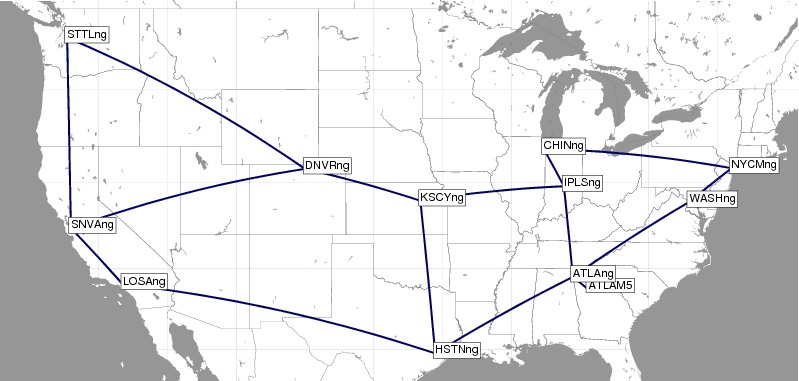} 
  \caption{Abilene network topology \cite{SNDlib} under study with 12 nodes and 15 links} \label{fig:abilene} \vspace{-0.3cm}
 \end{figure}

\begin{figure}
\center {
\includegraphics [width = 0.9\columnwidth] {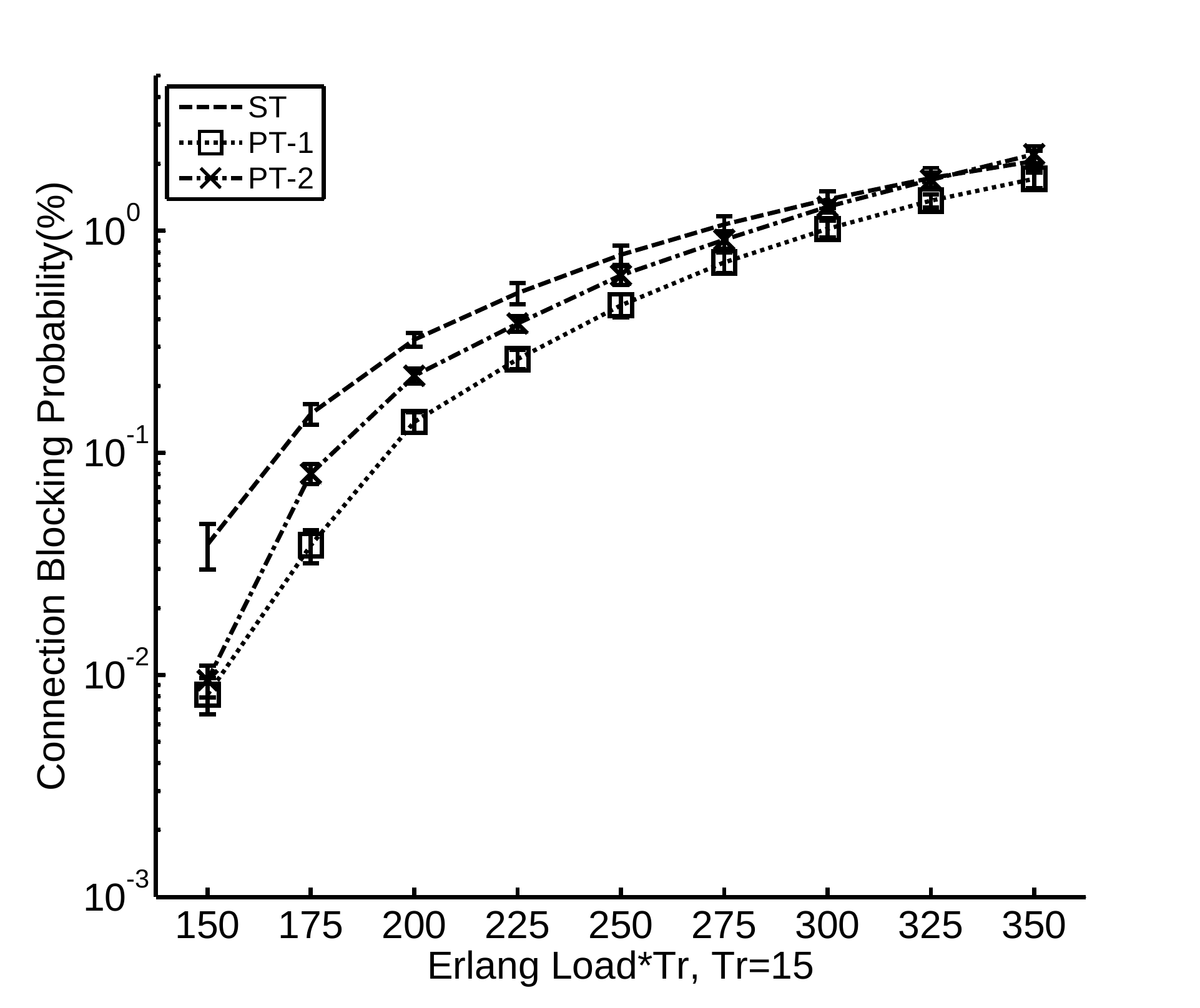}
} \caption{Blocking probability vs. network load, with no guard-band (GB=0) in Abilene network}\vspace{-0.3cm}
\label{fig:blockGb0}
\end{figure}

\begin{figure}
\center {
\includegraphics [width = 0.9\columnwidth] {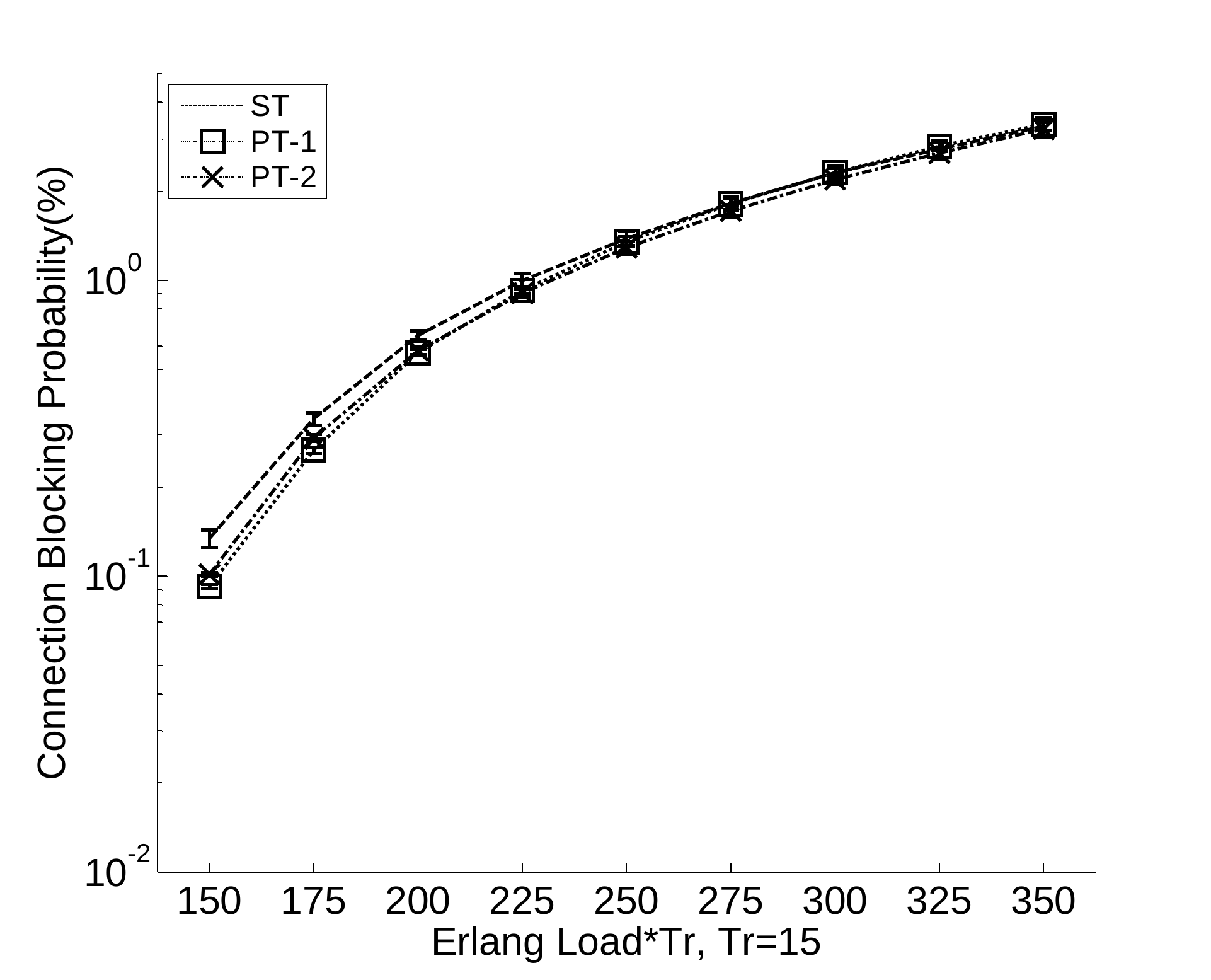}
} \caption{Blocking probability vs. network load, with a large guard-band (GB=3) in Abilene network}\vspace{-0.3cm}
\label{fig:blockGb3}
\end{figure}

  Finally, we evaluate the parallel transmission in a different network as shown in Fig.\ref{fig:abilene}. Fig.\ref{fig:blockGb0} and Fig.\ref{fig:blockGb3} show the blocking probability of connection requests requesting large number of sub-carriers ($T_r=15$). It can be seen that parallel transmission over multiple spectrum paths can reduce blocking probability in general in a small network. While the small network limits the maximum path length in PT-1, the large differential delay constraint leads to the better performance in terms of the blocking probability. As shown in Fig.\ref{fig:blockGb0}, PT-1 can still reduce around $0.4\%$ blocking probability comparing with ST at high network loads (e.g., $Load*T_r=350$), while PT-2 has almost the same performance as ST due to the strict differential delay constraint. 
 
 \section{Related Work}\label{relatedwork} 
\par Since 2010, IEEE 802.3ba has standardized  \emph{parallel transmission} as the solution  for high-speed Ethernet with  ever increasing transmission rate which can be beyond 100Gbps in the near future~\cite{802.3ba}.  In particular, 100GE has specified 10 parallel lanes, leveraging the mature 10Gbps technology in optical networks. To enable parallel transmission over Wide Area Network (WAN), the International Telecommunication Union-Telecommunication Standardization Sector (ITU-T), has augmented their G.709 OTN information structure for high-speed Ethernet at transmission of 40Gbps and 100Gbps~\cite{ITU-T:G.709}. 

\par It is no coincidence that about the same time, the optical network have started to approach the so-called 
\emph{optical capacity crunch}~\cite{Chralyvy:ecoc:2009}, where the capacities of single-mode fiber were shown to 
be reaching the Shannon limit  in a few years from now ~\cite{Winzer:2010}\cite{Winzer:2012}. 
Optical parallel transmission was coined in \cite{Winzer:2010} as the most promising  solution for the optical capacity crunch.  Optical spatial multiplexing and so-called  \emph{photonic MIMO (Multiple Input Multiple Output)} were proposed to explore the optical parallelism, 
  which can multiplex multiple fiber strands or multiplexing multiple modes in a fiber~\cite{Winzer:2012}.

\par In addition, optical networks are also shifting to  improving spectral efficiency.        It is known that conventional WDM networks have limit on the spectral efficiency. Therefore, it was easy for us to recognize that the inherent parallelism makes OFDM based optical networks   a valid candidate for the high-speed Ethernet parallel transmission.  Only by looking at early work \emph{SLICE} proposed in~\cite{Jinno:2009}, where optical spectrum is sliced into frequency slots, it can be seen that traffic is distributed to the parallel sub-carriers, each utilizes one frequency slot (in parallel). However, up to now, none of the existing work has addressed the issues of high-speed Ethernet parallel transmission in the context of elastic optical networks.  

\par In \cite{Chen:infocom:2013}, we presented a study of optimal algorithms for finding multiple parallel paths in elastic optical networks applicable to problems shown in this paper.   Also, optical parallel transmission  in support of high-speed Ethernet was investigated in our past paper \cite{Chen:JOCN:2012} for the first time, where an optimized parallel transmission framework based on OTN/WDM networks was proposed. In parallel to our work, proposals have been presented also in~\cite{Lu:Letter:2012} and~\cite{Zhu:JLT:2013}.    In \cite{Lu:Letter:2012}, the authors proposed a heuristic algorithm for dynamic multipath provisioning in OFDM based elastic optical networks. In \cite{Zhu:JLT:2013}, a hybrid single path and multipath routing scheme  was proposed for flexible online service provisioning in elastic optical networks. The authors proposed two heuristic algorithms, with consideration of path computation on the fly and usage of pre-computed paths.
\par The present paper is different from the previous body of work, since it is an optimization framework for dynamic computations of parallel transmission paths in elastic optical networks, with consideration of differential delay bounds within the spectrum bands and also among the spectrum band. Thus, the model is complete from the point of view of parallelization.
Moreover, this paper is entirely focused on optical parallel transmission to support high-speed Ethernet.  This present paper, thus, for the first time, investigates the effectiveness and feasibility of using parallel OFDM-basedoptical networks to support high-speed Ethernet according to the current IEEE and ITU-T standards, widely accepted by the industry.

\section{Conclusion}\label{conclusion}
\par Parallel transmission has been standardized in IEEE 802.3ba as the solution for high-speed Ethernet with every increasing transmission rates, which poses new requirements to the optical layer, which on its own, is reaching the Shannon limits on capacity. OFDM-based optical networks carry potential to support high-speed Ethernet due to its inherent parallelism. In this paper, we investigated the technical feasibility of parallel transmission in OFDM based networks to support high-speed Ethernet and designed a novel framework which is in-line with current IEEE and ITU-T standards. We formulated an optimization model based on Integer Linear Programming (ILP) for dynamic computation of multiple parallel paths and spectrum assignment, with consideration of differential delay issue caused by path diversity and fiber effects.  We also proposed an effective heuristic algorithm which can be applied for scenarios where the optimization model becomes intractable. The numerical results showed that differential delay issue of using multiple parallel paths is not an obstacle and parallel transmission in elastic optical networks is feasible. The parallel transmission framework proposed here can be used to effectively support high-speed Ethernet, while leveraging the maturity of low-speed optical technologies.

 \bibliographystyle{IEEEtran}
\bibliography{literature}

\end {document}